\documentclass{article} 
\usepackage{iclr2027_conference,times}

\usepackage{amsmath,amsfonts,bm}

\def\eqref#1{equation~\ref{#1}}

\def\1{\bm{1}}

\DeclareMathAlphabet{\mathsfit}{\encodingdefault}{\sfdefault}{m}{sl}
\SetMathAlphabet{\mathsfit}{bold}{\encodingdefault}{\sfdefault}{bx}{n}

\usepackage{hyperref}
\usepackage{url}
\usepackage{booktabs} 
\usepackage{multirow}
\usepackage{graphicx}
\usepackage{makecell}
\usepackage{algorithm}
\usepackage{algpseudocode}
\usepackage{wrapfig}
\usepackage{placeins}
\usepackage{float}
\usepackage{booktabs}
\usepackage{tabularx}
\usepackage{array}
\usepackage{placeins}
\usepackage{wrapfig}

\usepackage{caption}

\newcommand{\meanstdsmall}[2]{%
\begin{tabular}{@{}c@{}}
#1\\[-0.15em]
{\scriptsize $\pm$#2}
\end{tabular}%
}

\usepackage{xspace}
\newcommand{\mname}{PerturbRx\xspace}

\title{\mname: Learning Treatment-Conditioned Latent Transitions for Patient Drug Response Prediction}

\author{
Yoshitaka Inoue$^{1,4,5,*}$ \\
\texttt{inoue019@umn.edu} \\
\And
Minoh Jeong$^{2,*}$ \\
\texttt{minohj@inha.ac.kr} \\
\And
Alfred Hero$^{3}$ \\
\texttt{hero@umich.edu} \\
\And
Rui Kuang$^{1,\dagger}$ \\
\texttt{kuang@umn.edu} \\
\And
Augustin Luna$^{4,5,\dagger}$ \\
\texttt{augustin@nih.gov} \\
\\
$^{1}$Department of Computer Science, University of Minnesota, Minneapolis, MN, USA \\
$^{2}$Department of Electrical and Electronic Engineering, Inha University, Incheon, South Korea \\
$^{3}$Department of Electrical Engineering and Computer Science, University of Michigan, \\ Ann Arbor, MI, USA \\
$^{4}$Computational Biology Branch, National Library of Medicine, Bethesda, MD, USA \\
$^{5}$Therapeutic Development Branch, National Library of Medicine, Bethesda, MD, USA \\
\vspace{0.3em}
{\footnotesize
$^{*}$Equal contribution.
\quad
$^{\dagger}$Jointly supervised this work.
}
}

\iclrfinalcopy 
\begin{document}

\maketitle

\begin{abstract}
Scarce data and tumor heterogeneity limit patient-level cancer treatment-response prediction. Existing approaches predict response from pretreatment molecular profiles and drug representations, without explicitly modeling the molecular changes expected under treatment. We propose \mname, a treatment-conditioned representation learning framework that learns intervention-induced latent transitions and uses them as patient-drug response features. \mname trains a drug- and dose-conditioned transition predictor from context-matched but cell-unpaired control and treated single-cell populations, then freezes and transfers the predictor to pretreatment patient profiles without requiring post-treatment measurements. The transition is combined with patient and drug representations to predict response. Across TCGA and patient-derived xenograft benchmarks, \mname achieves the strongest aggregate predictive performance among the evaluated methods. These results support perturbation-pretrained latent transitions as useful representations for patient-level drug-response prediction.
\end{abstract}

\section{Introduction}

Patient-level cancer treatment-response prediction remains challenging because clinical response labels are scarce and tumors are molecularly heterogeneous~\citep{partin2023deep}. Many methods therefore transfer information from large-scale cancer cell-line screens~\citep{yang2012genomics, rees2016correlating}, despite substantial differences between cell lines and patient tumors~\citep{gillet2013clinical, ben2017patient}. Existing approaches focus primarily on transferable molecular representations, domain alignment, and supervised response adaptation~\citep{he2022context, zhang2026deepsadr}.

Most such approaches predict response from pretreated patient molecular profiles, cell-line molecular profiles, drug representations, or learned interactions among them. These features describe the patient, preclinical system, and therapy, but do not explicitly represent the molecular change expected under treatment. Treatment response may depend not only on the pretreatment state, but also on the direction and magnitude of the intervention-induced change.

Directly learning such changes from patients is difficult because matched pre- and post-treatment molecular profiles are rarely available at scale. Single-cell perturbation atlases instead provide context-matched control and treated populations across many drugs, doses, and cellular contexts~\citep{srivatsan2020massively, zhang2025tahoe}. Although individual cells are not paired before and after treatment, these data provide population-level evidence of intervention-induced state changes.

We introduce \mname, a treatment-aware representation learning framework that represents each patient-drug pair by a predicted drug-induced latent transition. \mname learns a drug- and dose-conditioned transition predictor from context-matched but unpaired control and treated single-cell populations in a frozen pretrained latent space. The pretrained predictor is then frozen and applied to a patient's pretreatment molecular profile and candidate drug to produce a patient- and drug-specific transition feature for response prediction. This formulation separates learning how treatments alter molecular states from learning which predicted alterations are associated with clinical response.

Our objective is not post-perturbation cell-state reconstruction. Rather, we ask whether intervention-pretrained latent transitions provide a more informative patient-drug representation than static baseline patient representation and drug features. We study three questions: whether intervention-conditioned transitions can be learned from unpaired perturbation populations, whether these transitions provide patient-response signal beyond static baseline representations, and how consistently they transfer across drugs.

We instantiate \mname using Tahoe-100M single-cell intervention profiles~\citep{zhang2025tahoe}, scFoundation expression embeddings~\citep{hao2024large}, and ChemBERTa molecular drug representations~\citep{chithrananda2020chemberta}. We evaluate it on TCGA treatment-response data~\citep{weinstein2013cancer} and an independent patient-derived xenograft (PDX) cohort~\citep{gao2015high} using source-stage alignment and held-out-drug analyses, controlled representation and predictor comparisons, and patient- and model-disjoint target evaluations. Our contributions are:
\begin{itemize}
\item We formulate patient drug-response prediction using predicted intervention-induced latent transitions and develop an inductive two-stage framework that can generate transition representations for previously unseen patient profiles and molecular compounds, without requiring post-treatment profiles in the target domain.

\item On a controlled held-out-patient TCGA benchmark, augmenting patient-drug features with pretrained transition features improves mean predictive performance over the corresponding static representation, a predicted post-intervention-state variant, transition-shuffling controls, and a random frozen-predictor control.

\item In an independent patient-derived xenograft cohort, we evaluate frozen transition transfer under model-disjoint evaluation and compare \mname with four established baselines on a shared 10-drug benchmark comprising 791 treatment episodes from 177 PDX models.
\end{itemize}

\begin{figure*}[!htbp]
\centering
\includegraphics[width=\textwidth]{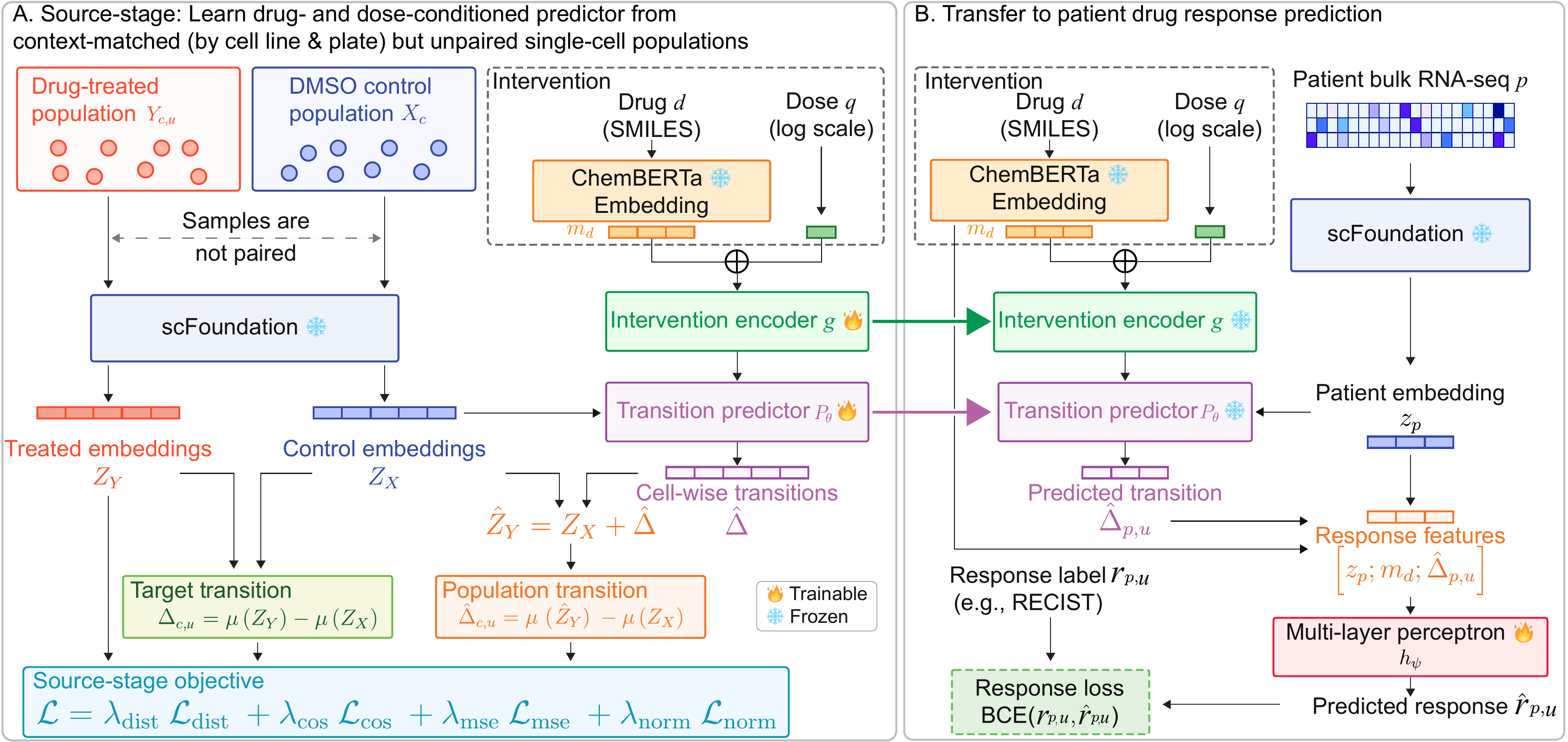}
\caption{
\textbf{Overview of population-level~\mname.} \textbf{(A)}~Source-stage learning from context-matched but unpaired control and treated single-cell populations. A drug- and dose-conditioned predictor maps control embeddings and intervention features to cell-wise transitions, which are added to the controls to form post-intervention embeddings. The predictor is trained using distributional, transition-direction, transition-displacement, and transition-magnitude objectives. \textbf{(B)}~Transfer of the transition predictor to patient drug-response prediction. The predictor generates an intervention-conditioned transition \(\hat{\Delta}_{p,u}\), and the response head predicts response from \([z_p; m_d; \hat{\Delta}_{p,u}]\).
} 
\label{fig:ia_jepa_overview}
\end{figure*}

\section{Related Work}

\paragraph{Patient Drug Response Prediction and Preclinical-to-Patient Transfer.}
Prior patient drug-response methods transfer information from cancer cell-line screens through approaches such as CODE-AE~\citep{he2022context} for domain adaptation, TransDRP~\citep{liu2025knowledge} for molecular-similarity-aware domain transfer, WISER~\citep{shubham2024wiser} for weak supervision, DeepSADR~\citep{zhang2026deepsadr} for supervised transfer, and GANDALF~\citep{jayagopal2025gandalf} for generative augmentation. These methods predict response from pretreatment patient profiles, drug representations, or learned interactions between them. In contrast, \mname transfers predicted treatment-induced state changes from single-cell perturbations to patient responses. Other treatment-aware approaches incorporate drug-genomic interactions, perturbation-derived representations, or simulated post-treatment states into response predictors~\citep{Rampasek2019drvae, sung2026deep, bang2026predictingtherapeuticoutcomealigning}. \mname instead learns a generative transition from matched control and treated populations and uses the predicted latent displacement as a response feature.

\paragraph{Single-Cell Perturbation and Latent Transition Modeling.}
Single-cell perturbation models such as scGen~\citep{lotfollahi2019scgen}, CPA~\citep{lotfollahi2023predicting}, CellOT~\citep{bunne2023learning}, PerturbDiff~\citep{yuan2026perturbdiff}, and Conditional Monge Gap (CMonge)~\citep{driessen2026conditional} learn intervention-induced changes from unpaired control and treated cell populations and are primarily evaluated by reconstruction of post-perturbation cellular states or distributions.  \mname extends this paradigm in both scale and task. At the source stage, it learns a drug- and context-conditioned transition model from a large-scale single-cell perturbation atlas spanning many drugs, doses, and cellular contexts. Rather than evaluating treated-state reconstruction, \mname transfers the learned drug-conditioned transition model to a patient-level target domain, where the predicted treatment-induced latent displacement is used for clinical drug-response prediction.

\section{Method}

We formulate \mname as a two-stage framework. In the source stage, the model learns drug- and dose-conditioned latent transition predictors from context-matched but unpaired control and treated single-cell populations. In the patient stage, the learned predictor is frozen and applied to patient molecular profiles to generate transition features for response prediction, as summarized in Figure~\ref{fig:ia_jepa_overview}. For clarity, we summarize the main notation used in the method in Appendix~\ref{app:notation}.

\subsection{Source and Target Formulation}

We consider context-matched but unpaired control and treated single-cell populations, where each expression profile is annotated with cell line, plate, drug condition, and dose. Dimethyl sulfoxide (DMSO) serves as the solvent control. Matched control and treated cells share the same context, whereas no individual cell is observed before and after treatment; supervision is therefore available only at the population level. A source context $c$ is defined by cell line $\ell$ and plate $b$,
\begin{align}
    c=(\ell,b),
\end{align}
and an intervention $u$ by drug \(d\) and dose \(q\),
\begin{align}
    u=(d,q).
\end{align}
For each condition \((\ell,b,d,q)\), we construct a context-matched control-treated population pair:
\begin{align}
    X_c 
    &= \{x_i : \mathrm{cell}(x_i)=\ell,\ \mathrm{plate}(x_i)=b,\ \mathrm{drug}(x_i)=\mathrm{DMSO}\}, \\
    Y_{c,u} 
    &= \{y_j : \mathrm{cell}(y_j)=\ell,\ \mathrm{plate}(y_j)=b,\ \mathrm{drug}(y_j)=d,\ \mathrm{dose}(y_j)=q\}.
\end{align}
Here, \(x_i\) and \(y_j\) denote single-cell gene-expression profiles. The set \(X_c=\{x_i\}_{i=1}^{n_c}\) denotes the DMSO control population for context \(c\), and \(Y_{c,u}=\{y_j\}_{j=1}^{m_{c,u}}\) denotes the population treated with intervention \(u\) in the same context. Their population sizes are \(n_c=|X_c|\) and \(m_{c,u}=|Y_{c,u}|\), respectively. We retain only conditions with sufficient control and treated cells. 

In the target domain, we observe patient response records
\begin{align}
\mathcal{D}_{\mathrm{tgt}}=\{(p,u,r_{p,u})\},
\end{align}
where $p$ is a pretreatment patient molecular profile, $u$ is a candidate treatment, and $r_{p,u}$ is a binary response label. Post-treatment molecular states are not observed. Our goal is to learn intervention-conditioned latent transition predictors from source perturbation populations and transfer them to patient response prediction.

\subsection{Intervention-Conditioned Latent Transition Predictor}

Let \(E\) be a frozen pretrained encoder that maps cells into a \(d\)-dimensional latent space. scFoundation produces a concatenated cell embedding with \(d=3072\). We embed the two populations as
\begin{align}
    Z_X = E(X_c) = \{z_i^X\}_{i=1}^{n_c}, \qquad
    Z_Y = E(Y_{c,u}) = \{z_j^Y\}_{j=1}^{m_{c,u}}.
\end{align}
The intervention is encoded as
\begin{align}
\alpha_u = g([m_d; \log q]),
\end{align}
where \(m_d\) is a frozen ChemBERTa embedding computed from the Simplified Molecular Input Line Entry System (SMILES) representation of drug \(d\), \(q\) is the perturbation dose, log-transformed to represent multiplicative dose differences, and \(g\) is a trainable intervention encoder. The predictor \(P_\theta:\mathbb{R}^{d}\times\mathbb{R}^{r}\rightarrow\mathbb{R}^{d}\) maps a latent cell-state embedding and an intervention embedding to a latent transition vector, where \(r=512\) denotes the intervention-embedding dimension.

\mname parameterizes an intervention-conditioned transition predictor with predictor $P_\theta$:
\begin{align}
\hat{\Delta}_i=P_\theta(z_i^X,\alpha_u),
\qquad
\hat{z}_i^Y=z_i^X+\hat{\Delta}_i.
\end{align}
The predicted treated population is $\hat{Z}_Y=\{\hat{z}_i^Y\}_{i=1}^{n_c}$. Because the data are unpaired, $\hat{z}_i^Y$ is not matched to a specific treated cell. Instead, $\hat{Z}_Y$ is trained to match $Z_Y$ distributionally while preserving the population-level transition:
\begin{align}
\Delta_{c,u}=\mu(Z_Y)-\mu(Z_X), \qquad
\hat{\Delta}_{c,u}=\mu(\hat{Z}_Y)-\mu(Z_X),
\end{align}
where $\mu(Z)=\frac{1}{|Z|}\sum_{z\in Z} z$. Architecture and optimization details for \(g\), \(P_\theta\), and the response head are provided in Table~\ref{tab:architecture_hyperparameters}.

\subsection{Population-Level Training Objective}

The source-stage objective combines distributional alignment via maximum mean discrepancy (MMD) with transition-direction, transition-displacement, and transition-magnitude constraints:
\begin{equation}
\label{eq:source_loss}
\begin{aligned}
\mathcal{L}_{\mathrm{src}}
={}&
\underbrace{
\mathrm{MMD}^2(\hat{Z}_Y,Z_Y)
}_{\text{distribution alignment}}
+
\lambda_{\mathrm{cos}}
\underbrace{
\left[
1-\cos\!\left(
\hat{\Delta}_{c,u},
\Delta_{c,u}
\right)
\right]
}_{\text{transition direction}}
\\
&+
\lambda_{\mathrm{mse}}
\underbrace{
\frac{1}{d}
\left\|
\hat{\Delta}_{c,u}
-
\Delta_{c,u}
\right\|_2^2
}_{\text{transition displacement}}
+
\lambda_{\mathrm{norm}}
\underbrace{
\left(
\left\|\hat{\Delta}_{c,u}\right\|_2
-
\left\|\Delta_{c,u}\right\|_2
\right)^2
}_{\text{transition magnitude}} .
\end{aligned}
\end{equation}

The distributional term aligns the predicted and observed treated populations, while the delta-aware terms constrain the predicted population transition relative to the empirical transition. We use \(\lambda_{\mathrm{cos}}=0.1\), \(\lambda_{\mathrm{mse}}=3.0\), and \(\lambda_{\mathrm{norm}}=10^{-3}\), selected using source-stage validation only. The empirical MMD estimator, kernel bandwidths, and numerical details are provided in Appendix~\ref{app:source_objective_details}, and the loss-weight selection procedure is described in Appendix~\ref{app:source_loss_selection}.

\subsection{Frozen Predictor Transfer to Patients}

After source training, the transition predictor defines an intervention-conditioned predictor
\begin{equation}
\label{eq:patient_transition}
\hat{\Delta}_{p,u}
=
P_\theta(z_p,\alpha_u),
\qquad
\hat{z}_{p,u}
=
z_p+\hat{\Delta}_{p,u},
\end{equation}
where $z_p = E_{\mathrm{pat}}(p)$ is the scFoundation embedding of the pretreatment patient profile and $\alpha_u$ is the intervention embedding. The frozen scFoundation pipeline is applied to gene-harmonized patient bulk RNA-seq profiles, yielding embeddings in the source-cell $d=3072$-dimensional output space without source-target alignment. Because the predictor operates on representations rather than fixed identities, it can be applied to unseen patients and drugs. Here, ``patient-specific transition'' denotes the patient-conditioned output of the source-trained predictor. Because matched patient pre- and post-treatment profiles are unavailable, it does not represent a validated molecular transition.

We use a frozen predictor: the source-stage intervention encoder and transition predictor are fixed, and only a response head is trained on response labels. In the downstream experiment, the response head $h_{\psi}$ uses the static patient-drug representation augmented with the predicted transition:
{
\setlength{\abovedisplayskip}{4pt}
\setlength{\belowdisplayskip}{4pt}
\setlength{\abovedisplayshortskip}{4pt}
\setlength{\belowdisplayshortskip}{4pt}
\begin{equation}
\hat{r}_{p,u}
=
h_{\psi}\!\left([z_p; m_d; \hat{\Delta}_{p,u}]\right).
\end{equation}
}

The response head is trained with binary cross-entropy,
$\mathcal{L}_{\mathrm{resp}}
=
\mathrm{BCE}(r_{p,u},\hat{r}_{p,u}).$
We additionally evaluate alternative response features, including the pretreatment patient embedding \(z_p\), the predicted post-intervention state \(\hat z_{p,u}\), and concatenated feature variants, as representation ablations. Since patient post-treatment molecular states are unavailable, transferred transitions are evaluated through response prediction and predictor-specific controls.

\section{Experiments}

\subsection{Experimental Setup}
\label{sec:experimental_setup}

We pretrain \mname on Tahoe-100M using frozen scFoundation embeddings and ChemBERTa drug representations. We then transfer the frozen intervention encoder and transition predictor to three target benchmarks. TCGA-186 is the primary held-out-patient benchmark, with 186 patient-drug records across five drugs and retrospective relapse-time proxy labels (Appendix Table~\ref{tab:appendix_patient_drug_stats}); its drug-dependent cancer-type composition is summarized in Appendix Table~\ref{tab:appendix_patient_cancer_stats}. TCGA-508 provides a larger patient-grouped evaluation with 508 treatment episodes from 462 patients across five drugs (Appendix Table~\ref{tab:expanded_drug_distribution}). The independent PDX benchmark provides cross-domain preclinical evaluation; the comparison uses 791 treatment episodes from 177 PDX models across the 10 drugs shared by all methods. Unlike the TCGA cohorts, which were constructed from GDC gene-level read counts, the PDX cohort uses published FPKM-quantified bulk RNA-seq profiles; both domains are mapped through the frozen scFoundation representation pipeline before response modeling. Cohort construction, expression preprocessing, and response-label definitions are provided in Appendix~\ref{app:data_and_cohorts}.

TCGA-186 uses drug-stratified held-out-patient splits, TCGA-508 uses patient-grouped splits, and PDX uses five model-disjoint outer folds. For \mname, the source-trained intervention encoder and transition predictor remain frozen and only the response head is trained. All methods share the same held-out partitions and are trained or adapted using non-held-out data according to their method-specific procedures. Evaluation protocols and implementation details are provided in Appendix~\ref{app:training_and_evaluation}. We report AUROC and AUPRC. Because clinical and in vivo treatment doses cannot be directly mapped to the Tahoe-100M micromolar scale, inference dose is treated as a representation setting rather than calibrated exposure; dose sensitivity is reported in Appendices~\ref{app:additional_results}, \ref{app:tcga508_dose_sensitivity}, and \ref{app:pdx_shared_dose_sensitivity}.

\subsection{Source-Stage Perturbation Alignment}
\label{sec:source_alignment}

\begin{figure}[t]
\centering

\includegraphics[width=\textwidth]{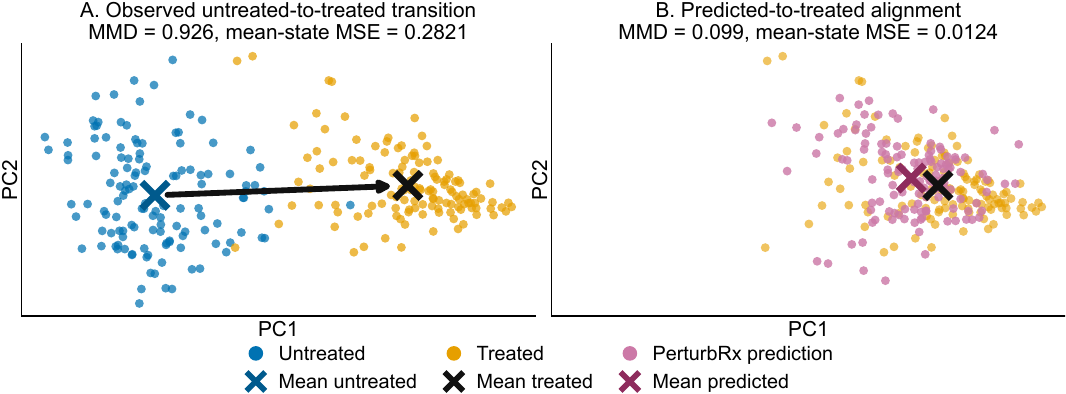}


{\small
\textbf{C. Quantitative comparison on randomly held-out conditions}


\setlength{\tabcolsep}{9pt}
\renewcommand{\arraystretch}{1.08}

\begin{tabular}{lccc}
\toprule
\textbf{Method}
& \textbf{MMD $\downarrow$}
& \textbf{Transition cosine $\uparrow$}
& \textbf{Delta MSE $\downarrow$} \\
\midrule

Identity
& 0.1082
& -
& 0.0212 \\

Global mean transition
& 0.1053
& 0.2185
& 0.0204 \\

Linear regression
& 0.0899
& 0.4759
& 0.0154 \\

MLP
& $0.0801^{\dagger}$
& $0.5686^{\dagger}$
& $0.0122^{\dagger}$ \\

Conditional AE
& \underline{0.0758}
& \underline{0.6476}
& \underline{0.0108} \\

\textbf{\mname}
& \textbf{0.0682}
& \textbf{0.6715}
& \textbf{0.0096} \\

\bottomrule
\end{tabular}
}
\caption{
\textbf{Source-stage perturbation prediction.}
\textbf{A.} Observed untreated-to-treated shift for an example held-out condition.
\textbf{B.} Alignment between predicted and observed treated states for the same condition.
Crosses denote population means.
\textbf{C.} Performance across 800 randomly held-out conditions.
\textbf{Bold}, \underline{underline}, and $\ensuremath{^{\dagger}}$ indicate the best, second-best, and third-best results, respectively.
}
\label{fig:source_validation}

\end{figure}

We first evaluate whether \mname learns intervention-conditioned transition predictors in the source perturbation domain. For each matched context-intervention pair $(c,u)$, we compare the predicted treated latent population $\hat{Z}_Y$ with the observed treated latent population $Z_Y$, and the predicted population transition $\hat{\Delta}_{c,u}$ with the empirical transition $\Delta_{c,u}$.

Figure~\ref{fig:source_validation} summarizes source-stage prediction and controlled baseline comparisons. On 800 random held-out conditions, \mname achieved an MMD of 0.0682, a transition cosine of 0.6715, and a delta MSE of 0.0096, outperforming all evaluated baselines across all three metrics. Performance improved progressively from the global mean transition to linear regression, a nonlinear MLP, and a state-conditioned conditional autoencoder. The conditional autoencoder achieved an MMD of 0.0758, a transition cosine of 0.6476, and a delta MSE of 0.0108, while \mname further improved each metric. The progressive gains across increasingly expressive and state-aware baselines suggest that both intervention conditioning and explicit transition modeling contribute to source-stage performance. These results indicate that the source-stage gains are not explained solely by nonlinear drug-dose mapping or by conditioning on the initial cellular state, and support the additional value of cell-wise state-conditioned transition prediction.

\subsection{Transition Representations and Patient Response Prediction}
\label{sec:patient_response_prediction}

We next evaluate whether predicted intervention-induced transitions provide response-predictive information beyond static patient and drug representations. We first compare \mname with existing patient drug-response methods on the TCGA-186 benchmark and then use controlled representation ablations and negative controls to isolate the contribution of the pretrained transition representation.

\begin{table*}[!b]
\centering
\caption{Patient-response prediction on the TCGA-186 benchmark.}
\label{tab:patient_transfer_drugwise}

\scriptsize
\setlength{\tabcolsep}{3.0pt}
\renewcommand{\arraystretch}{1.05}

\resizebox{\textwidth}{!}{%
\begin{tabular}{lcccccccccccc}
\toprule
& \multicolumn{2}{c}{\textbf{Overall}}
& \multicolumn{2}{c}{\textbf{Cisplatin}}
& \multicolumn{2}{c}{\textbf{Fluorouracil}$^{\ddagger}$}
& \multicolumn{2}{c}{\textbf{Gemcitabine}$^{\ddagger}$}
& \multicolumn{2}{c}{\textbf{Sorafenib}}
& \multicolumn{2}{c}{\textbf{Temozolomide}} \\
\cmidrule(lr){2-3}
\cmidrule(lr){4-5}
\cmidrule(lr){6-7}
\cmidrule(lr){8-9}
\cmidrule(lr){10-11}
\cmidrule(lr){12-13}

\textbf{Methods}
& \textbf{AUROC$\uparrow$} & \textbf{AUPRC$\uparrow$}
& \textbf{AUROC$\uparrow$} & \textbf{AUPRC$\uparrow$}
& \textbf{AUROC$\uparrow$} & \textbf{AUPRC$\uparrow$}
& \textbf{AUROC$\uparrow$} & \textbf{AUPRC$\uparrow$}
& \textbf{AUROC$\uparrow$} & \textbf{AUPRC$\uparrow$}
& \textbf{AUROC$\uparrow$} & \textbf{AUPRC$\uparrow$} \\
\midrule

CODE-AE
& \meanstdsmall{0.534\ensuremath{^{\dagger}}}{0.053}
& \meanstdsmall{\underline{0.564}}{0.051}
& \meanstdsmall{0.609\ensuremath{^{\dagger}}}{0.073}
& \meanstdsmall{0.666\ensuremath{^{\dagger}}}{0.058}
& \meanstdsmall{0.548\ensuremath{^{\dagger}}}{0.147}
& \meanstdsmall{0.516}{0.134}
& \meanstdsmall{0.360}{0.023}
& \meanstdsmall{0.413}{0.023}
& \meanstdsmall{0.631\ensuremath{^{\dagger}}}{0.015}
& \meanstdsmall{0.689\ensuremath{^{\dagger}}}{0.040}
& \meanstdsmall{0.581\ensuremath{^{\dagger}}}{0.028}
& \meanstdsmall{0.587\ensuremath{^{\dagger}}}{0.031} \\

DeepSADR
& \meanstdsmall{0.530}{0.011}
& \meanstdsmall{0.519}{0.027}
& \meanstdsmall{0.569}{0.103}
& \meanstdsmall{0.596}{0.093}
& \textbf{\meanstdsmall{0.749}{0.113}}
& \textbf{\meanstdsmall{0.768}{0.137}}
& \meanstdsmall{0.573\ensuremath{^{\dagger}}}{0.122}
& \textbf{\meanstdsmall{0.601}{0.121}}
& \meanstdsmall{\underline{0.720}}{0.126}
& \meanstdsmall{\underline{0.712}}{0.148}
& \meanstdsmall{0.508}{0.111}
& \meanstdsmall{0.528}{0.106} \\

TransDRP
& \meanstdsmall{0.494}{0.061}
& \meanstdsmall{0.500}{0.068}
& \meanstdsmall{0.484}{0.176}
& \meanstdsmall{0.598}{0.104}
& \meanstdsmall{0.354}{0.094}
& \meanstdsmall{0.391}{0.044}
& \textbf{\meanstdsmall{0.593}{0.068}}
& \meanstdsmall{\underline{0.569}}{0.070}
& \meanstdsmall{0.473}{0.220}
& \meanstdsmall{0.581}{0.156}
& \meanstdsmall{0.496}{0.217}
& \meanstdsmall{0.552}{0.175} \\

WISER
& \meanstdsmall{\underline{0.580}}{0.053}
& \meanstdsmall{0.563\ensuremath{^{\dagger}}}{0.062}
& \textbf{\meanstdsmall{0.709}{0.151}}
& \textbf{\meanstdsmall{0.756}{0.141}}
& \meanstdsmall{0.480}{0.168}
& \meanstdsmall{0.526\ensuremath{^{\dagger}}}{0.175}
& \meanstdsmall{\underline{0.578}}{0.076}
& \meanstdsmall{0.547\ensuremath{^{\dagger}}}{0.081}
& \meanstdsmall{0.467}{0.105}
& \meanstdsmall{0.574}{0.115}
& \meanstdsmall{\underline{0.589}}{0.124}
& \meanstdsmall{\underline{0.621}}{0.118} \\

\midrule

\mname
& \textbf{\meanstdsmall{0.626}{0.019}}
& \textbf{\meanstdsmall{0.600}{0.029}}
& \meanstdsmall{\underline{0.684}}{0.056}
& \meanstdsmall{\underline{0.733}}{0.093}
& \meanstdsmall{\underline{0.674}}{0.104}
& \meanstdsmall{\underline{0.645}}{0.134}
& \meanstdsmall{0.551}{0.082}
& \meanstdsmall{0.530}{0.055}
& \textbf{\meanstdsmall{0.753}{0.213}}
& \textbf{\meanstdsmall{0.801}{0.174}}
& \textbf{\meanstdsmall{0.653}{0.142}}
& \textbf{\meanstdsmall{0.680}{0.107}} \\

\bottomrule
\end{tabular}%
}

\vspace{1.5mm}
\begin{minipage}{0.98\textwidth}
\scriptsize
\textit{Note:}
Mean $\pm$ standard deviation across five patient-grouped evaluations. Bold, underline, and $\ensuremath{^{\dagger}}$ denote the best, second-best, and third-best results; $^\ddagger$ denotes exact Tahoe-100M support. Drug-specific AUPRC should be interpreted relative to positive-class prevalence.
\end{minipage}
\end{table*}

Table~\ref{tab:patient_transfer_drugwise} reports aggregate and drug-stratified performance. \mname achieved the highest aggregate performance, with an AUROC of \(0.626 \pm 0.019\) and an AUPRC of \(0.600 \pm 0.029\). The strongest comparator achieved an AUROC of \(0.580 \pm 0.053\) (WISER) and an AUPRC of \(0.564 \pm 0.051\) (CODE-AE). Performance varied across drugs. \mname achieved the highest mean AUROC and AUPRC for Sorafenib and Temozolomide and the second-highest values for Cisplatin and Fluorouracil, whereas other methods performed better for Gemcitabine. Thus, the aggregate improvement does not imply uniformly superior performance for every drug.

\begin{table*}[!htbp]


\noindent
\begin{minipage}[t]{0.47\textwidth}
\vspace{0pt}

\caption{Ablation results on TCGA-186.}
\label{tab:tcga_ablation}

\vspace{0.5mm}

\scriptsize
\setlength{\tabcolsep}{3.0pt}
\renewcommand{\arraystretch}{1.03}

\begin{tabularx}{\linewidth}{@{}Xcc@{}}
\toprule
\textbf{Input representation / control}
& \textbf{AUROC$\uparrow$}
& \textbf{AUPRC$\uparrow$} \\
\midrule

Patient embedding
& \meanstdsmall{0.547}{0.040}
& \meanstdsmall{0.518}{0.062} \\

Patient + drug embedding
& \meanstdsmall{0.581}{0.019}
& \meanstdsmall{0.556}{0.055} \\

Patient + drug + shuffled transition
& \meanstdsmall{0.566}{0.029}
& \meanstdsmall{0.551}{0.048} \\

Patient + drug + random transition
& \meanstdsmall{0.589\ensuremath{^{\dagger}}}{0.021}
& \meanstdsmall{0.574\ensuremath{^{\dagger}}}{0.039} \\

Patient + drug + predicted post-state
& \meanstdsmall{\underline{0.609}}{0.051}
& \meanstdsmall{\underline{0.582}}{0.062} \\

\textbf{\mname\ (Patient + drug + transition)}
& \textbf{\meanstdsmall{0.626}{0.019}}
& \textbf{\meanstdsmall{0.600}{0.029}} \\

\bottomrule
\end{tabularx}

\vspace{0.5mm}

\parbox{\linewidth}{%
\tiny
Results are mean $\pm$ standard deviation over five held-out-patient splits at \(0.05\,\mu\mathrm{M}\). The shuffled-transition control permutes transitions among patients receiving the same drug. The random control uses the same architecture without pretraining. 
}

\end{minipage}
\hfill
\begin{minipage}[t]{0.5\textwidth}
\vspace{0pt}

We next tested whether the aggregate improvement was specifically associated with the pretrained transition representation. All variants in Table~\ref{tab:tcga_ablation} use the same MLP response head, evaluation splits, patient-label supervision budget, and \(0.05\,\mu\mathrm{M}\) inference dose. The pretreatment patient embedding alone achieved \(0.547 \pm 0.040\) AUROC and \(0.518 \pm 0.062\) AUPRC. Adding the ChemBERTa drug representation improved performance to \(0.581 \pm 0.019\) AUROC and \(0.556 \pm 0.055\) AUPRC, while augmenting this static patient-drug representation with the pretrained \mname transition further increased performance to \(0.626 \pm 0.019\) AUROC and \(0.600 \pm 0.029\) AUPRC.

\end{minipage}


\vspace{1.5mm}

\begin{minipage}{\textwidth}
\normalsize

The controlled variants did not reproduce this improvement. Replacing the transition with the predicted post-intervention state yielded \(0.609 \pm 0.051\) AUROC and \(0.582 \pm 0.062\) AUPRC, suggesting that the predicted displacement is more informative than the corresponding post-treatment state. A random frozen predictor achieved \(0.589 \pm 0.021\) AUROC and \(0.574 \pm 0.039\) AUPRC, arguing against output dimensionality, predictor architecture, or random nonlinear feature expansion alone as explanations for the gain. Shuffling pretrained transitions among patients receiving the same drug reduced performance to \(0.566 \pm 0.029\) AUROC and \(0.551 \pm 0.048\) AUPRC. Together, these controls indicate that source-stage perturbation pretraining and patient-transition correspondence contribute to downstream prediction. They do not establish, however, that the predicted transitions reproduce molecular changes observed after treatment in individual patients.

\end{minipage}

\end{table*}

\subsection{TCGA-508 Patient-Grouped Evaluation}

\begin{table*}[!b]
\centering
\caption{Patient-response prediction on the TCGA-508 patient-grouped benchmark.}
\label{tab:tcga508_results}

\scriptsize
\setlength{\tabcolsep}{3.0pt}
\renewcommand{\arraystretch}{1.05}

\resizebox{\textwidth}{!}{%
\begin{tabular}{lcc|cc|cc|cc|cc|cc}
\toprule

& \multicolumn{2}{c}{\textbf{Overall}}
& \multicolumn{2}{c}{\textbf{Cisplatin}}
& \multicolumn{2}{c}{\textbf{Docetaxel}$^{\ddagger}$}
& \multicolumn{2}{c}{\textbf{Fluorouracil}$^{\ddagger}$}
& \multicolumn{2}{c}{\textbf{Gemcitabine}$^{\ddagger}$}
& \multicolumn{2}{c}{\textbf{Paclitaxel}$^{\ddagger}$} \\

\cmidrule(lr){2-3}
\cmidrule(lr){4-5}
\cmidrule(lr){6-7}
\cmidrule(lr){8-9}
\cmidrule(lr){10-11}
\cmidrule(lr){12-13}

\textbf{Methods}
& \textbf{AUROC$\uparrow$} & \textbf{AUPRC$\uparrow$}
& \textbf{AUROC$\uparrow$} & \textbf{AUPRC$\uparrow$}
& \textbf{AUROC$\uparrow$} & \textbf{AUPRC$\uparrow$}
& \textbf{AUROC$\uparrow$} & \textbf{AUPRC$\uparrow$}
& \textbf{AUROC$\uparrow$} & \textbf{AUPRC$\uparrow$}
& \textbf{AUROC$\uparrow$} & \textbf{AUPRC$\uparrow$} \\

\midrule

CODE-AE
& \meanstdsmall{0.525\ensuremath{^{\dagger}}}{0.061}
& \meanstdsmall{0.649}{0.193}
& \meanstdsmall{0.600\ensuremath{^{\dagger}}}{0.028}
& \meanstdsmall{0.841\ensuremath{^{\dagger}}}{0.027}
& \meanstdsmall{0.453}{0.056}
& \meanstdsmall{0.557}{0.038}
& \meanstdsmall{0.545}{0.024}
& \meanstdsmall{0.683}{0.042}
& \meanstdsmall{0.472\ensuremath{^{\dagger}}}{0.044}
& \meanstdsmall{0.366}{0.043}
& \meanstdsmall{\underline{0.556}}{0.064}
& \meanstdsmall{\underline{0.799}}{0.040} \\

DeepSADR
& \meanstdsmall{\underline{0.565}}{0.147}
& \meanstdsmall{\underline{0.735}}{0.093}
& \meanstdsmall{0.511}{0.071}
& \meanstdsmall{0.807}{0.121}
& \textbf{\meanstdsmall{0.724}{0.244}}
& \textbf{\meanstdsmall{0.812}{0.160}}
& \meanstdsmall{\underline{0.565}}{0.103}
& \meanstdsmall{\underline{0.741}}{0.072}
& \textbf{\meanstdsmall{0.675}{0.055}}
& \textbf{\meanstdsmall{0.582}{0.234}}
& \meanstdsmall{0.351}{0.093}
& \meanstdsmall{0.732\ensuremath{^{\dagger}}}{0.258} \\

TransDRP
& \meanstdsmall{0.508}{0.089}
& \meanstdsmall{0.665\ensuremath{^{\dagger}}}{0.174}
& \meanstdsmall{\underline{0.613}}{0.164}
& \textbf{\meanstdsmall{0.885}{0.054}}
& \meanstdsmall{0.425}{0.241}
& \meanstdsmall{0.588}{0.197}
& \textbf{\meanstdsmall{0.587}{0.192}}
& \meanstdsmall{0.710\ensuremath{^{\dagger}}}{0.178}
& \meanstdsmall{0.422}{0.083}
& \meanstdsmall{0.418\ensuremath{^{\dagger}}}{0.079}
& \meanstdsmall{0.493}{0.253}
& \meanstdsmall{0.726}{0.122} \\

WISER
& \meanstdsmall{0.507}{0.016}
& \meanstdsmall{0.650}{0.151}
& \meanstdsmall{0.480}{0.040}
& \meanstdsmall{0.829}{0.017}
& \meanstdsmall{0.515\ensuremath{^{\dagger}}}{0.074}
& \meanstdsmall{0.607\ensuremath{^{\dagger}}}{0.055}
& \meanstdsmall{0.511}{0.032}
& \meanstdsmall{0.667}{0.034}
& \meanstdsmall{\underline{0.512}}{0.032}
& \meanstdsmall{\underline{0.422}}{0.031}
& \meanstdsmall{0.519\ensuremath{^{\dagger}}}{0.051}
& \meanstdsmall{0.724}{0.033} \\

\midrule

\mname
& \textbf{\meanstdsmall{0.692}{0.055}}
& \textbf{\meanstdsmall{0.787}{0.055}}
& \textbf{\meanstdsmall{0.631}{0.168}}
& \meanstdsmall{\underline{0.864}}{0.085}
& \meanstdsmall{\underline{0.711}}{0.141}
& \meanstdsmall{\underline{0.802}}{0.112}
& \meanstdsmall{0.562\ensuremath{^{\dagger}}}{0.116}
& \textbf{\meanstdsmall{0.748}{0.098}}
& \meanstdsmall{0.472\ensuremath{^{\dagger}}}{0.095}
& \meanstdsmall{0.395}{0.052}
& \textbf{\meanstdsmall{0.715}{0.135}}
& \textbf{\meanstdsmall{0.842}{0.081}} \\

\bottomrule
\end{tabular}%
}

\vspace{0.8mm}

\begin{minipage}{0.98\textwidth}
\scriptsize
\textit{Note:}
Results are mean $\pm$ standard deviation across five patient-grouped evaluations. Best, second-best, and third-best results are indicated by \textbf{bold}, \underline{underline}, and $\ensuremath{^{\dagger}}$, respectively. $^{\ddagger}$ indicates exact Tahoe-100M support by normalized drug name or canonicalized molecular structure. Drug-specific AUPRC values should be interpreted relative to the corresponding positive-class prevalence.
\end{minipage}

\end{table*}

We further evaluated \mname\ on TCGA-508, comprising 508 treatment episodes from 462 patients across five drugs. Five patient-grouped splits assign all episodes from each patient to the same fold, evaluating generalization to unseen patients rather than drug-cold transfer. Cohort construction, response definitions, and drug composition are provided in Appendix~\ref{app:expanded_tcga_cohort}.

Table~\ref{tab:tcga508_results} reports aggregate and drug-stratified performance. \mname achieved the highest overall AUROC (\(0.692 \pm 0.055\)) and AUPRC (\(0.787 \pm 0.055\)), but performance varied across drugs, with other methods performing best for several treatments. Thus, aggregate performance does not imply uniform superiority across treatments. Drug-specific AUPRC should be interpreted relative to positive-class prevalence, which varies substantially across drugs (Appendix Table~\ref{tab:expanded_drug_distribution}). 


\begin{figure*}[t]

\noindent
\begin{minipage}[t]{0.47\textwidth}
\vspace{0pt}

\centering
\includegraphics[width=\linewidth]{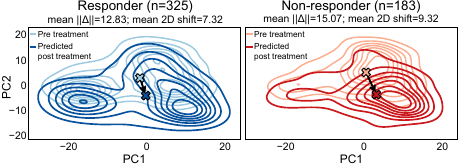}

\vspace{0.5mm}

\caption{
\textbf{Predicted transition geometry in TCGA-508.} PCA was fit jointly to pretreatment and predicted post-intervention embeddings for all 508 treatment episodes. Contours show the two state distributions within each response group, and black arrows connect their group centroids. Panel annotations report sample size, mean predicted transition magnitude \(\|\hat{\Delta}\|_2\), and mean episode-level displacement in the two-dimensional PCA projection. The reported 2D shift is the mean paired episode-level displacement, not the centroid-arrow length.
}
\label{fig:tcga508_transition_geometry}

\end{minipage}
\hfill
\begin{minipage}[t]{0.50\textwidth}
\vspace{0pt}

To characterize the transferred transition geometry, we examined predicted latent displacements in TCGA-508 (Figure~\ref{fig:tcga508_transition_geometry}). Responders exhibited smaller mean predicted transition magnitudes than non-responders (\(12.83\) vs.\ \(15.07\)), with corresponding mean PCA displacements of \(7.32\) and \(9.32\), respectively. 

After adjustment for drug and cancer type with patient-clustered standard errors, this difference remained significant (\(\beta_{\mathrm{R-NR}}=-1.56\), 95\% CI \([-2.64,-0.49]\), \(p=0.004\)) and was robust to alternative handling of sparsely represented cancer types. 

Exploratory analyses across TCGA-186, TCGA-508, and PDX showed substantial treatment- and cohort-level heterogeneity (Appendix~\ref{app:transition_magnitude}), indicating that the TCGA-508 pattern should not be interpreted as a universal transition signature. These results concern \emph{predicted} latent post-intervention states rather than measured molecular changes.

\end{minipage}

\end{figure*}

\begin{table*}[t]


\noindent
\begin{minipage}[t]{0.47\textwidth}
\vspace{0pt}

\caption{Ablation results on the TCGA-508.}
\label{tab:tcga508_ablation}

\vspace{0.5mm}

\scriptsize
\setlength{\tabcolsep}{3.0pt}
\renewcommand{\arraystretch}{1.03}

\begin{tabularx}{\linewidth}{@{}Xcc@{}}
\toprule
\textbf{Input representation / control}
& \textbf{AUROC$\uparrow$}
& \textbf{AUPRC$\uparrow$} \\
\midrule

Patient embedding
& \meanstdsmall{0.638}{0.038}
& \meanstdsmall{0.756}{0.041} \\

Patient + drug embedding
& \meanstdsmall{0.681\ensuremath{^{\dagger}}}{0.046}
& \meanstdsmall{0.780\ensuremath{^{\dagger}}}{0.055} \\

Patient + drug + shuffled transition
& \meanstdsmall{0.665}{0.030}
& \meanstdsmall{0.770}{0.038} \\

Patient + drug + random transition
& \meanstdsmall{0.676}{0.046}
& \meanstdsmall{0.773}{0.046} \\

Patient + drug + predicted post-state
& \meanstdsmall{\underline{0.685}}{0.048}
& \meanstdsmall{\underline{0.781}}{0.054} \\

\textbf{\mname\ (Patient + drug + transition)}
& \textbf{\meanstdsmall{0.692}{0.055}}
& \textbf{\meanstdsmall{0.787}{0.055}} \\

\bottomrule
\end{tabularx}

\vspace{0.5mm}

\parbox{\linewidth}{%
\tiny
Results are mean $\pm$ standard deviation across five patient-grouped evaluation folds.
All variants use the same response-head architecture, training procedure, and response-label supervision.
The random control uses the same transition architecture without pretraining.
}

\end{minipage}
\hfill
\begin{minipage}[t]{0.5\textwidth}
\vspace{0pt}

We next tested whether the aggregate improvement on TCGA-508 was
specifically associated with the pretrained transition representation.
All variants in Table~\ref{tab:tcga508_ablation} use the same response
head, patient-grouped folds, training procedure, and response-label
supervision.
The pretreatment patient embedding achieved
\(0.638 \pm 0.038\) AUROC and \(0.756 \pm 0.041\) AUPRC.
Adding the ChemBERTa drug representation improved performance to
\(0.681 \pm 0.046\) AUROC and \(0.780 \pm 0.055\) AUPRC, while
augmenting this static patient-drug representation with the pretrained
\mname transition further increased performance to
\(0.692 \pm 0.055\) AUROC and \(0.787 \pm 0.055\) AUPRC.

\end{minipage}


\vspace{1.5mm}

\begin{minipage}{\textwidth}
\normalsize

The controlled variants again failed to reproduce the full improvement.
Replacing the transition with the predicted post-intervention state
yielded \(0.685 \pm 0.048\) AUROC and \(0.781 \pm 0.054\) AUPRC.
A random frozen transition predictor achieved
\(0.676 \pm 0.046\) AUROC and \(0.773 \pm 0.046\) AUPRC, while
shuffling the pretrained transition reduced performance to
\(0.665 \pm 0.030\) AUROC and \(0.770 \pm 0.038\) AUPRC.
Together with the TCGA-186 ablations, these results show that the
transition-specific advantage is reproduced in a larger patient-grouped
cohort and is not explained solely by additional feature dimensionality
or by adding a predicted post-treatment state.

\end{minipage}

\end{table*}

\subsection{PDX Model-Disjoint Evaluation}
\label{sec:pdx_evaluation}

\begin{table*}[!t]
\centering
\caption{
Drug-stratified performance on the shared 10-drug PDX benchmark. Values are mean $\pm$ standard deviation across five model-disjoint outer folds. Best, second-best, and third-best results are indicated by \textbf{bold}, \underline{underline}, and $\dagger$, respectively. Drug-specific AUPRC should be interpreted relative to the corresponding positive-class prevalence. $^{\ddagger}$ indicates exact Tahoe-100M support by normalized drug name or canonicalized molecular structure.
}
\label{tab:pdx_drugwise}

\scriptsize
\setlength{\tabcolsep}{2.8pt}
\renewcommand{\arraystretch}{1.03}

\textbf{AUROC}

\vspace{0.4mm}

\resizebox{\textwidth}{!}{%
\begin{tabular}{lccccccccccc}
\toprule
Method
& Overall
& Alpelisib$^{\ddagger}$
& Buparlisib
& Erlotinib$^{\ddagger}$
& Fluorouracil$^{\ddagger}$
& Gemcitabine$^{\ddagger}$
& Paclitaxel$^{\ddagger}$
& Ribociclib$^{\ddagger}$
& Ruxolitinib
& Tamoxifen
& Trametinib$^{\ddagger}$ \\
\midrule

CODE-AE
& \meanstdsmall{0.432}{0.035}
& \meanstdsmall{0.472}{0.193}
& \meanstdsmall{0.474}{0.106}
& \meanstdsmall{\underline{0.400}}{0.253}
& \meanstdsmall{0.445}{0.153}
& \meanstdsmall{\underline{0.467}}{0.317}
& \meanstdsmall{\underline{0.589}}{0.140}
& \meanstdsmall{0.486}{0.047}
& \meanstdsmall{0.439}{0.103}
& \meanstdsmall{0.476}{0.268}
& \meanstdsmall{0.407}{0.245} \\

DeepSADR
& \meanstdsmall{\underline{0.637}}{0.024}
& \meanstdsmall{\underline{0.587}}{0.079}
& \textbf{\meanstdsmall{0.595}{0.074}}
& \meanstdsmall{0.300}{0.217}
& \meanstdsmall{0.304}{0.196}
& \meanstdsmall{0.462}{0.349}
& \meanstdsmall{0.511}{0.065}
& \meanstdsmall{0.497}{0.082}
& \textbf{\meanstdsmall{0.762}{0.158}}
& \textbf{\meanstdsmall{0.719}{0.206}}
& \meanstdsmall{0.430}{0.258} \\

TransDRP
& \meanstdsmall{0.470}{0.026}
& \meanstdsmall{0.418}{0.124}
& \meanstdsmall{0.438}{0.107}
& \meanstdsmall{0.367\ensuremath{^{\dagger}}}{0.183}
& \meanstdsmall{0.508\ensuremath{^{\dagger}}}{0.373}
& \meanstdsmall{0.463}{0.359}
& \meanstdsmall{0.540}{0.232}
& \meanstdsmall{0.525\ensuremath{^{\dagger}}}{0.047}
& \meanstdsmall{0.342}{0.133}
& \meanstdsmall{0.181}{0.262}
& \meanstdsmall{\underline{0.640}}{0.123} \\

WISER
& \meanstdsmall{0.492\ensuremath{^{\dagger}}}{0.031}
& \meanstdsmall{0.527\ensuremath{^{\dagger}}}{0.047}
& \meanstdsmall{0.503\ensuremath{^{\dagger}}}{0.145}
& \meanstdsmall{\underline{0.400}}{0.384}
& \meanstdsmall{\underline{0.616}}{0.138}
& \meanstdsmall{0.465\ensuremath{^{\dagger}}}{0.214}
& \meanstdsmall{0.554\ensuremath{^{\dagger}}}{0.174}
& \meanstdsmall{\underline{0.576}}{0.084}
& \meanstdsmall{0.575\ensuremath{^{\dagger}}}{0.209}
& \meanstdsmall{0.595\ensuremath{^{\dagger}}}{0.337}
& \meanstdsmall{0.537\ensuremath{^{\dagger}}}{0.322} \\

\midrule

\mname
& \textbf{\meanstdsmall{0.669}{0.025}}
& \textbf{\meanstdsmall{0.631}{0.109}}
& \meanstdsmall{\underline{0.544}}{0.054}
& \textbf{\meanstdsmall{0.767}{0.149}}
& \textbf{\meanstdsmall{0.680}{0.172}}
& \textbf{\meanstdsmall{0.473}{0.334}}
& \textbf{\meanstdsmall{0.634}{0.264}}
& \textbf{\meanstdsmall{0.607}{0.072}}
& \meanstdsmall{\underline{0.682}}{0.221}
& \meanstdsmall{\underline{0.657}}{0.211}
& \textbf{\meanstdsmall{0.760}{0.189}} \\

\bottomrule
\end{tabular}%
}

\vspace{1.5mm}

\textbf{AUPRC}

\vspace{0.4mm}

\resizebox{\textwidth}{!}{%
\begin{tabular}{lccccccccccc}
\toprule
Method
& Overall
& Alpelisib$^{\ddagger}$
& Buparlisib
& Erlotinib$^{\ddagger}$
& Fluorouracil$^{\ddagger}$
& Gemcitabine$^{\ddagger}$
& Paclitaxel$^{\ddagger}$
& Ribociclib$^{\ddagger}$
& Ruxolitinib
& Tamoxifen
& Trametinib$^{\ddagger}$ \\
\midrule

CODE-AE
& \meanstdsmall{0.389}{0.024}
& \meanstdsmall{0.499}{0.196}
& \meanstdsmall{0.477}{0.092}
& \meanstdsmall{0.454\ensuremath{^{\dagger}}}{0.237}
& \meanstdsmall{0.524}{0.165}
& \meanstdsmall{0.652}{0.241}
& \meanstdsmall{0.517\ensuremath{^{\dagger}}}{0.177}
& \meanstdsmall{0.396}{0.061}
& \meanstdsmall{0.122}{0.020}
& \meanstdsmall{0.133}{0.068}
& \meanstdsmall{0.650}{0.191} \\

DeepSADR
& \meanstdsmall{\underline{0.554}}{0.018}
& \meanstdsmall{\underline{0.561}}{0.078}
& \textbf{\meanstdsmall{0.624}{0.100}}
& \meanstdsmall{0.320}{0.093}
& \meanstdsmall{0.389}{0.069}
& \meanstdsmall{0.672\ensuremath{^{\dagger}}}{0.247}
& \meanstdsmall{0.500}{0.086}
& \meanstdsmall{0.435}{0.111}
& \textbf{\meanstdsmall{0.488}{0.266}}
& \textbf{\meanstdsmall{0.333}{0.376}}
& \meanstdsmall{0.721}{0.116} \\

TransDRP
& \meanstdsmall{0.410}{0.034}
& \meanstdsmall{0.444}{0.087}
& \meanstdsmall{0.487}{0.083}
& \meanstdsmall{\underline{0.497}}{0.208}
& \meanstdsmall{0.635\ensuremath{^{\dagger}}}{0.260}
& \textbf{\meanstdsmall{0.726}{0.230}}
& \meanstdsmall{0.508}{0.081}
& \meanstdsmall{\underline{0.481}}{0.139}
& \meanstdsmall{0.170}{0.050}
& \meanstdsmall{0.169}{0.055}
& \meanstdsmall{\underline{0.852}}{0.068} \\

WISER
& \meanstdsmall{0.446\ensuremath{^{\dagger}}}{0.036}
& \meanstdsmall{0.530\ensuremath{^{\dagger}}}{0.094}
& \meanstdsmall{0.544\ensuremath{^{\dagger}}}{0.163}
& \meanstdsmall{0.449}{0.312}
& \meanstdsmall{\underline{0.679}}{0.134}
& \meanstdsmall{\underline{0.703}}{0.145}
& \meanstdsmall{\underline{0.531}}{0.204}
& \meanstdsmall{0.465\ensuremath{^{\dagger}}}{0.102}
& \meanstdsmall{0.253\ensuremath{^{\dagger}}}{0.220}
& \meanstdsmall{\underline{0.309}}{0.392}
& \meanstdsmall{0.778\ensuremath{^{\dagger}}}{0.171} \\

\midrule

\mname
& \textbf{\meanstdsmall{0.608}{0.037}}
& \textbf{\meanstdsmall{0.590}{0.089}}
& \meanstdsmall{\underline{0.559}}{0.056}
& \textbf{\meanstdsmall{0.776}{0.036}}
& \textbf{\meanstdsmall{0.749}{0.128}}
& \meanstdsmall{0.646}{0.254}
& \textbf{\meanstdsmall{0.617}{0.294}}
& \textbf{\meanstdsmall{0.535}{0.069}}
& \meanstdsmall{\underline{0.361}}{0.268}
& \meanstdsmall{0.308\ensuremath{^{\dagger}}}{0.387}
& \textbf{\meanstdsmall{0.914}{0.058}} \\

\bottomrule
\end{tabular}%
}

\end{table*}

We next evaluated whether the predictive advantage of \mname\ transfers to an independent preclinical domain. The PDX benchmark contains 791 treatment episodes from 177 models across 10 drugs. Evaluation used five model-disjoint outer folds, with all episodes from the same model assigned to the same fold, thereby testing generalization to unseen models. Cohort construction and the full evaluation protocol are described in Appendix~\ref{app:pdx_cohort} and Appendix~\ref{app:pdx_protocol},
respectively.

On the shared 10-drug benchmark, \mname\ achieved the highest aggregate performance among the evaluated methods, with an AUROC of $0.669 \pm 0.025$ and an AUPRC of $0.608 \pm 0.037$. The strongest comparator, DeepSADR, achieved $0.637 \pm 0.024$ AUROC and $0.554 \pm 0.018$ AUPRC. These results indicate that the aggregate advantage observed in the TCGA benchmarks also extends to an independent PDX domain under model-disjoint evaluation.

Table~\ref{tab:pdx_drugwise} reports drug-stratified performance. Relative performance varied across treatments. \mname\ achieved the highest mean AUROC for Alpelisib, Erlotinib, Fluorouracil, Gemcitabine, Paclitaxel, Ribociclib, and Trametinib, while DeepSADR performed best for Buparlisib, Ruxolitinib, and Tamoxifen. A similar pattern was observed for AUPRC: \mname\ achieved the highest mean for Alpelisib, Erlotinib, Fluorouracil, Paclitaxel, Ribociclib, and Trametinib, whereas other methods performed best for the remaining drugs. Thus, the aggregate improvement again does not imply uniform superiority across individual treatments. Because drug-specific response prevalence and fold-level sample sizes vary substantially, the drug-stratified results should be interpreted primarily as evidence of treatment-specific heterogeneity rather than definitive rankings for individual drugs.

Across the three target benchmarks, the incremental utility of the predicted transition varied substantially across drugs and was not consistently explained by nearest-source chemical similarity. In PDX, the largest transition-related degradations occurred in the two treatment strata with the lowest observed responder prevalence, with detailed sensitivity analyses provided in Appendix~\ref{app:transition_utility_reliability}.

\section{Conclusion and Discussion}

We introduced \mname, a treatment-conditioned representation learning framework that learns drug- and dose-conditioned latent transition predictors from context-matched but unpaired control and treated single-cell populations and transfers the resulting displacements to treatment-response prediction. Across TCGA-186, TCGA-508, and PDX, augmenting patient-drug representations with pretrained transition features improved aggregate performance and outperformed representation and predictor controls. The gains were reproduced across distinct target cohorts, although performance remained heterogeneous across treatments. These results support perturbation-pretrained latent transitions as useful treatment-response representations beyond static patient and drug features.

Several limitations remain. The target benchmarks are retrospective and heterogeneous. TCGA-186 uses relapse-time proxy labels rather than prospective clinical-response endpoints, and treatment is partially confounded with cancer type. The source and target domains also differ: Tahoe-100M contains single-cell profiles from cultured cell lines, whereas the target cohorts contain bulk tumor RNA-seq. Thus, the shared scFoundation representation should not be interpreted as establishing domain alignment or biological equivalence. The source-stage minimum cell-count requirement may exclude strongly cytotoxic conditions with substantial treatment-induced cell depletion.

Drug-specific transition utility was heterogeneous and was not consistently explained by source-drug chemical similarity across target benchmarks (Appendix~\ref{app:transition_utility_reliability}). Although TCGA-508 showed an exploratory positive association between nearest-source chemical similarity and transition benefit, this pattern did not replicate in TCGA-186 or PDX. Exact Tahoe-100M support was also not necessary for successful transfer, as illustrated by Sorafenib, which lacked an exact source match but had Regorafenib as a close molecular neighbor. These findings suggest that source chemical coverage alone is insufficient to predict when transferred transition features will improve response prediction.

Although \mname can be applied to unseen patients and drugs, unseen-drug generalization is evaluated at the source-transition level. Transfer to drugs with limited or no source-stage perturbation support therefore remains a limitation, and inference dose is treated as a representation setting rather than a calibrated clinical or in vivo exposure. The predicted transitions are also not directly validated against matched post-treatment profiles or molecular mechanisms. Finally, \mname benefits from large-scale perturbation-atlas pretraining unavailable to the evaluated baselines, although random-predictor and transition-shuffling controls partially address this asymmetry.

Future work should validate predicted transitions using prospective cohorts with matched pre- and post-treatment molecular measurements. Broader perturbation coverage may improve transfer to drugs with limited source support, while mechanistic interpretation and dose mapping could strengthen biological relevance. Determining what makes predicted transitions useful for individual drugs remains an important open question.

\clearpage

\section*{Acknowledgments}

This research was supported in part by the Intramural Research Program of the National Institutes of Health (NIH) (ZIALM240126). The contributions of the NIH author(s) are considered Works of the United States Government. The findings and conclusions presented in this paper are those of the author(s) and do not necessarily reflect the views of the NIH or the U.S. Department of Health and Human Services.

\section*{AI Use Statement}

Generative AI tools were used to assist with language editing, code debugging, and document formatting. All scientific claims, experimental design decisions, implementations, analyses, and reported results were reviewed and verified by the authors. The authors take full responsibility for the content of the paper.

\section*{Ethics Statement}

This study uses publicly available, de-identified molecular and treatment-response datasets, including TCGA and PDX. No new human subjects were recruited, and no identifiable patient information was accessed. The analyses were conducted for methodological research and are not intended to support direct clinical decision-making. Predicted latent transitions and treatment-response scores should not be interpreted as validated clinical recommendations.

\section*{Reproducibility Statement}
We document dataset access, cohort construction, preprocessing, architectures, training objectives, hyperparameters, random seeds, and evaluation protocols in the appendices. The source and target datasets are publicly available, with exact access locations and preprocessing versions reported in the corresponding dataset sections. Code and processed metadata will be made publicly available upon publication, with links to the public GitHub repository and an archived Zenodo release included in the final version.

\bibliography{iclr2027_conference}
\bibliographystyle{iclr2027_conference}

\appendix

\section{Notation}
\label{app:notation}

Table~\ref{tab:notation} summarizes the notation specific to the
source-stage transition model and its transfer to patient response prediction.

\begin{table}[t]
\centering
\caption{Notation used in the \mname formulation.}
\label{tab:notation}
\small
\setlength{\tabcolsep}{6pt}
\renewcommand{\arraystretch}{1.05}
\begin{tabular}{ll}
\toprule
\textbf{Symbol} & \textbf{Meaning} \\
\midrule

\(c=(\ell,b)\)
& Source context defined by cell line \(\ell\) and plate \(b\) \\

\(u=(d,q)\)
& Intervention defined by drug \(d\) and dose \(q\) \\

\(X_c\)
& DMSO control population in context \(c\) \\

\(Y_{c,u}\)
& Treated population in context \(c\) under intervention \(u\) \\

\(Z_X, Z_Y\)
& Frozen-encoder embeddings of \(X_c\) and \(Y_{c,u}\) \\

\(\alpha_u\)
& Learned intervention embedding for \(u\) \\

\(P_\theta\)
& Intervention-conditioned latent transition predictor \\

\(\hat{Z}_Y\)
& Predicted treated latent population \\

\(\Delta_{c,u}\)
& Empirical population-level transition \\

\(\hat{\Delta}_{c,u}\)
& Predicted population-level transition \\

\(z_p\)
& Pretreatment patient embedding \\

\(\hat{\Delta}_{p,u}\)
& Predicted patient- and intervention-conditioned transition feature \\

\(\hat{z}_{p,u}\)
& Predicted post-intervention patient embedding \\

\(h_\psi\)
& Target-domain response head \\

\bottomrule
\end{tabular}
\end{table}

\section{Data and Cohort Construction}
\label{app:data_and_cohorts}

\subsection{Tahoe-100M Source Dataset}
\label{app:tahoe_source}

We use Tahoe-100M as the source-stage single-cell perturbation dataset. Context-matched but unpaired control and treated cell populations were constructed by pairing DMSO-treated control cells with drug-treated cells from the same cell line and plate. Each source condition is indexed by cell line, plate, drug, and dose. Table~\ref{tab:tahoe_source_dataset_statistics} summarizes the resulting source dataset.

\begin{table}[!htbp]
\centering
\caption{
Statistics of the Tahoe-100M source-stage perturbation dataset.
Observed treated conditions are defined as observed cell line, plate, drug, and dose combinations.
Matched DMSO groups are defined by cell line and plate.
}
\label{tab:tahoe_source_dataset_statistics}
\begin{tabular}{lr}
\toprule
Statistic & Value \\
\midrule
Post-QC cells & 100.6M \\
Plates & 14 \\
Cell lines & 50 \\
Treated drugs & 379 \\
Dose levels & 0.05, 0.5, 5.0 $\mu$M \\
Observed treated conditions & 65{,}218 \\
Median cells per treated condition & 1{,}210 \\
Matched DMSO groups & $\sim$700 \\
Median DMSO cells per group & 2{,}811 \\
\bottomrule
\end{tabular}
\end{table}

\subsection{TCGA-186 Patient Benchmark}
\label{app:patient_preprocessing}

We use a TCGA subset of the processed WISER-style patient drug response benchmark~\citep{shubham2024wiser} for target-stage evaluation. We retain patient-drug records whose patient identifier is a TCGA barcode and whose drug is one of five chemotherapy agents: Fluorouracil, Temozolomide, Gemcitabine, Cisplatin, or Sorafenib. Following the processed WISER-style benchmark, patient response labels are defined from cancer relapse time after chemotherapy: records with relapse times greater than the benchmark-defined median are labeled as responders, whereas records with relapse times less than the median are labeled as non-responders. Thus, the target labels are binary relapse-time proxy labels derived from processed treatment-response and relapse-time annotations, rather than prospective RECIST clinical response endpoints.

We obtained TCGA bulk RNA-seq expression data from the National Cancer Institute Genomic Data Commons (GDC)~\citep{heath2021nci}. Specifically, we used the gene-level read-count files (\texttt{data\_mrna\_seq\_read\_counts.txt}) distributed for the TCGA-GDC study collections in the TCGA treatment-response data repository. The expression files were processed on a per-study basis and combined at the sample-by-gene level. TCGA sample barcodes were reduced to patient-level barcodes using the first 12 characters of each barcode (e.g., \texttt{TCGA-XX-YYYY-01A} $\rightarrow$ \texttt{TCGA-XX-YYYY}). For patients with multiple RNA-seq samples, we selected a single representative sample using a predefined sample-type priority that favors primary tumor samples over recurrent, metastatic, blood-derived normal, and solid-tissue-normal samples. The priority order was primary solid tumor (01), primary blood-derived cancer--peripheral blood (03), primary blood-derived cancer--bone marrow (09), recurrent solid tumor (02), metastatic (06), blood-derived normal (10), and solid tissue normal (11). This selection procedure follows the expression-matrix construction used for the GDC TCGA treatment-response cohorts. 

After sample selection, the resulting gene-level count matrix was used as input to the target-stage expression preprocessing pipeline. Counts were subjected to library-size normalization followed by $\log(1+x)$ transformation and then encoded using the frozen scFoundation model, yielding a 3072-dimensional patient representation. The same GDC-derived read-count source, patient-level sample-selection procedure, normalization, and scFoundation embedding pipeline were used for the TCGA-508 cohort, ensuring that both TCGA benchmarks were represented in a common target-domain embedding space.

The TCGA treatment-response repository constructs its treatment cohorts from cBioPortal-style TCGA study directories containing treatment timelines and associated molecular data. The underlying treatment-response pipeline standardizes treatment outcomes to CR, PR, SD, and PD, normalizes therapeutic-agent names, consolidates treatment records into episodes, and identifies single-agent treatment episodes for downstream modeling. 

After matching the response records to the GDC-derived expression profiles, the TCGA-186 benchmark contains 186 patient-drug records from 186 unique patients across the five eligible drugs. The benchmark includes 91 responders and 95 non-responders. Table~\ref{tab:appendix_patient_drug_stats} summarizes the drug-wise response composition of the benchmark, and Table~\ref{tab:appendix_patient_cancer_stats} reports the corresponding distribution across TCGA cancer types and treatments.

\begin{table}[!htbp]
\centering
\caption{
Drug-level composition of TCGA-186. Response labels are relapse-time proxies derived from the processed WISER-style benchmark.
}
\label{tab:appendix_patient_drug_stats}
\begin{tabular}{lrrr}
\toprule
Drug & Samples & Responders & Non-responders \\
\midrule
Cisplatin     & 40  & 20 & 20 \\
Fluorouracil & 28  & 12 & 16 \\
Gemcitabine  & 46  & 23 & 23 \\
Sorafenib    & 26  & 13 & 13 \\
Temozolomide & 46  & 23 & 23 \\
\midrule
Total         & 186 & 91 & 95 \\
\bottomrule
\end{tabular}
\end{table}

\begin{table}[t]
\centering
\caption{
Cancer-type and drug composition of TCGA-186. Entries denote the number of unique patients associated with each cancer type and treatment. Cancer types were assigned using TCGA project metadata from the Genomic Data Commons.
}
\label{tab:appendix_patient_cancer_stats}

\setlength{\tabcolsep}{3.5pt}
\renewcommand{\arraystretch}{0.95}

\begin{tabular}{lrrrrrr}
\toprule
Cancer type
& Cis.
& Flu.
& Gem.
& Sor.
& Tem.
& Total \\
\midrule
Adrenocortical       & 0  & 0  & 0  & 1  & 0  & 1  \\
Bladder              & 0  & 0  & 4  & 0  & 0  & 4  \\
Brain                & 0  & 0  & 0  & 0  & 42 & 42 \\
Cervical             & 5  & 0  & 0  & 0  & 0  & 5  \\
Colorectal           & 0  & 8  & 0  & 0  & 0  & 8  \\
Esophagus/Stomach    & 1  & 20 & 0  & 0  & 0  & 21 \\
Gallbladder          & 0  & 0  & 3  & 0  & 0  & 3  \\
Head and Neck        & 22 & 0  & 0  & 0  & 0  & 22 \\
Kidney               & 0  & 0  & 1  & 6  & 0  & 7  \\
Liver                & 0  & 0  & 2  & 19 & 0  & 21 \\
Lung                 & 2  & 0  & 3  & 0  & 1  & 6  \\
Melanoma             & 0  & 0  & 0  & 0  & 3  & 3  \\
Pancreas             & 0  & 0  & 30 & 0  & 0  & 30 \\
Prostate             & 0  & 0  & 1  & 0  & 0  & 1  \\
Sarcoma              & 1  & 0  & 2  & 0  & 0  & 3  \\
Uterine              & 9  & 0  & 0  & 0  & 0  & 9  \\
\midrule
Total                & 40 & 28 & 46 & 26 & 46 & 186 \\
\bottomrule
\end{tabular}
\end{table}

\subsection{TCGA-508 Treatment-Response Cohort}
\label{app:expanded_tcga_cohort}

\paragraph{Treatment records and response labels.}
We constructed the TCGA-508 cohort from treatment timeline records in the TCGA Genomic Data Commons (GDC)~\citep{heath2021nci}. We retained pharmacologic treatment records with a non-empty therapeutic-agent annotation and a response that could be mapped to complete response (CR), partial response (PR), stable disease (SD), or progressive disease (PD). The source annotation ``No Response'' was mapped to PD. We then defined an objective-response-rate (ORR)-like binary endpoint as
\[
y_{\mathrm{ORR}} =
\begin{cases}
1, & \mathrm{CR\ or\ PR},\\
0, & \mathrm{SD\ or\ PD}.
\end{cases}
\]

\paragraph{Episode construction and drug eligibility.}
Therapeutic-agent names were normalized, with predefined salt forms and naming variants mapped to a common active-ingredient name. Treatment records were consolidated into episodes using the TCGA dataset, patient identifiers, treatment dates, treatment types, and response annotations. When both treatment dates were missing, the corresponding source records were retained as separate episodes. We restricted the analysis to single-agent episodes.

Candidate drugs were required to have at least 30 single-agent episodes, at least 10 ORR-positive episodes, at least 10 ORR-negative episodes, representation in at least three TCGA cancer types, and no single cancer type accounting for more than 80\% of episodes. These criteria yielded five eligible drugs: Cisplatin, Docetaxel, Fluorouracil, Gemcitabine, and Paclitaxel.

\begin{table}[!htbp]
\centering
\caption{
Drug-level composition of the TCGA-508 analytic cohort after expression matching.
ORR-positive denotes complete or partial response, whereas ORR-negative denotes stable or progressive disease.
An exact Tahoe match indicates agreement under harmonized drug-name and canonical-SMILES matching.
}
\label{tab:expanded_drug_distribution}
\small
\setlength{\tabcolsep}{4pt}
\begin{tabular}{lccccc}
\toprule
Drug
& Episodes
& ORR+
& ORR-
& Prevalence
& Tahoe match \\
\midrule
Cisplatin     & 151 & 120 & 31 & 0.795 & No  \\
Docetaxel     & 52  & 29  & 23 & 0.558 & Yes \\
Fluorouracil  & 106 & 69  & 37 & 0.651 & Yes \\
Gemcitabine   & 112 & 42  & 70 & 0.375 & Yes \\
Paclitaxel    & 87  & 65  & 22 & 0.747 & Yes \\
\midrule
Total         & 508 & 325 & 183 & 0.640 & - \\
\bottomrule
\end{tabular}
\end{table}

\paragraph{Expression matching and final benchmark.}
Treatment episodes were matched to pretreatment TCGA bulk RNA-seq profiles obtained from the GDC. We used the GDC gene-level read-count profiles as the expression source and applied the same frozen scFoundation patient-embedding pipeline used for TCGA-186. Sample-level TCGA barcodes were mapped to patient-level barcodes, and a single representative pretreatment sample was selected for each patient using the common sample-selection procedure. The resulting count profiles were subjected to library-size normalization and $\log(1+x)$ transformation before frozen scFoundation encoding, yielding 3072-dimensional patient embeddings. Thus, TCGA-186 and TCGA-508 were processed in the same target-domain expression space and with the same frozen patient-embedding pipeline.

To maintain a common evaluation cohort across \mname and all baseline methods, we retained only treatment episodes for which the required pretreatment expression profile, drug representation, response label, and shared preprocessing outputs were available. The resulting benchmark comprised 508 treatment episodes from 462 unique patients across five drugs and 24 cancer types. Multiple treatment episodes from the same patient were retained; accordingly, all evaluation splits were grouped by patient identifier to prevent patient overlap between training and test sets.

The final cohort contained 325 ORR-positive and 183 ORR-negative episodes, corresponding to a positive-class prevalence of 0.640. Table~\ref{tab:expanded_drug_distribution} summarizes the drug-level composition, and Table~\ref{tab:tcga508_tumor_composition} shows the distribution of treatment episodes across tumor types and drugs.

\begin{table}[t]
\centering
\caption{
Tumor-type composition of the TCGA-508 benchmark across the five evaluated drugs.
Entries denote treatment-episode counts for each tumor type-drug combination.
}
\label{tab:tcga508_tumor_composition}
\small
\setlength{\tabcolsep}{5pt}
\resizebox{\textwidth}{!}{
\begin{tabular}{lrrrrrr}
\toprule
\textbf{Tumor type}
& \textbf{Cisplatin}
& \textbf{Docetaxel}
& \textbf{Fluorouracil}
& \textbf{Gemcitabine}
& \textbf{Paclitaxel}
& \textbf{Total} \\
\midrule
BRCA   & 0  & 25 & 0  & 4  & 48 & 77 \\
PAAD   & 1  & 0  & 10 & 66 & 0  & 77 \\
CESC   & 62 & 0  & 4  & 1  & 2  & 69 \\
STAD   & 5  & 2  & 51 & 0  & 1  & 59 \\
HNSC   & 44 & 3  & 0  & 1  & 2  & 50 \\
BLCA   & 10 & 3  & 0  & 17 & 1  & 31 \\
UCEC   & 5  & 2  & 0  & 1  & 14 & 22 \\
COAD   & 0  & 0  & 20 & 0  & 0  & 20 \\
LUAD   & 6  & 7  & 0  & 2  & 3  & 18 \\
UCS    & 2  & 3  & 0  & 1  & 10 & 16 \\
READ   & 0  & 0  & 16 & 0  & 0  & 16 \\
LUSC   & 3  & 3  & 0  & 5  & 3  & 14 \\
PLMESO & 4  & 0  & 0  & 3  & 0  & 7 \\
ESCA   & 0  & 0  & 5  & 1  & 1  & 7 \\
CHOL   & 1  & 0  & 0  & 4  & 0  & 5 \\
HGSOC  & 1  & 1  & 0  & 1  & 1  & 4 \\
PRAD   & 0  & 3  & 0  & 0  & 0  & 3 \\
SKCM   & 1  & 0  & 0  & 1  & 1  & 3 \\
NSGCT  & 2  & 0  & 0  & 0  & 0  & 2 \\
HCC    & 0  & 0  & 0  & 2  & 0  & 2 \\
PRCC   & 1  & 0  & 0  & 1  & 0  & 2 \\
THYM   & 2  & 0  & 0  & 0  & 0  & 2 \\
ACC    & 1  & 0  & 0  & 0  & 0  & 1 \\
MNET   & 0  & 0  & 0  & 1  & 0  & 1 \\
\midrule
\textbf{Total}
& \textbf{151}
& \textbf{52}
& \textbf{106}
& \textbf{112}
& \textbf{87}
& \textbf{508} \\
\bottomrule
\end{tabular}
}
\end{table}

\subsection{PDX Cohort Construction}
\label{app:pdx_cohort}

We constructed an independent patient-derived xenograft (PDX) treatment-response benchmark from the NIBR-PDXE resource~\citep{gao2015high} by matching pretreatment bulk tumor RNA-seq profiles to treatment-response records at the PDX-model level. Individual PDX models could contribute multiple treatment episodes corresponding to different drugs.

\paragraph{Response-label construction.}
Treatment response was derived from longitudinal tumor-volume measurements and the associated categorical response annotations~\citep{gao2015high,nguyen2021predicting}. To obtain a consistent endpoint, we recomputed the response category for each evaluable model--treatment episode from the reported \textit{BestResponse} and \textit{BestAvgResponse} measurements using the NIBR-PDXE response criteria. Complete response (CR) required $\mathrm{BestResponse}<-95$ and $\mathrm{BestAvgResponse}<-40$; partial response (PR) required $\mathrm{BestResponse}<-50$ and $\mathrm{BestAvgResponse}<-20$; stable disease (SD) required $\mathrm{BestResponse}<35$ and $\mathrm{BestAvgResponse}<30$; and all remaining evaluable episodes were classified as progressive disease (PD). Categories were assigned hierarchically in the order CR, PR, SD, and PD. The recomputed CR/PR/SD/PD base categories agreed with the corresponding base categories in the provided response annotations for all evaluable records. We then grouped CR, PR, and SD as responders and PD as non-response:
\[
y_{\mathrm{PDX}} =
\begin{cases}
1, & \text{CR, PR, or SD},\\
0, & \text{PD}.
\end{cases}
\]
Records without sufficient information to assign one of these four response categories were excluded.

\paragraph{Expression data and molecular preprocessing.}
Pretreatment bulk RNA-seq profiles were obtained from the published NIBR-PDXE expression resource accompanying the original study~\citep{gao2015high}. Unlike the TCGA target cohorts, which were constructed from GDC gene-level read counts, the PDX source matrix provides precomputed gene-level RNA-seq expression quantified as FPKM. The source matrix, originally organized as genes by PDX models, was transposed to a model-by-gene representation, and duplicated gene symbols were collapsed by averaging their expression values.

For compatibility with the frozen scFoundation encoder, PDX expression profiles were aligned to the 19{,}264-gene scFoundation vocabulary, with genes absent from the PDX expression matrix assigned a value of zero. Each model-level expression profile was then normalized to a total expression of \(10{,}000\) and transformed using \(\log(1+x)\). The resulting profiles were encoded using the same frozen scFoundation model used for the TCGA target cohorts, yielding a \(3072\)-dimensional pretreatment representation for each PDX model. Thus, the downstream normalization and frozen representation model were shared across the PDX and TCGA target evaluations, while the upstream expression quantification differed: the PDX cohort used published FPKM values, whereas the TCGA cohorts were constructed from GDC gene-level read counts.

Drug names were harmonized to canonical treatment identities. Retained small molecules were matched to valid chemical structures and represented using the frozen \(768\)-dimensional ChemBERTa encoder used throughout the target-stage evaluations. Episodes lacking a matched expression profile, valid response label, or valid molecular representation were excluded.

\paragraph{Shared 10-drug benchmark.}
For cross-method comparison, we restricted evaluation to the 10 drugs shared across all evaluated methods: Fluorouracil, Buparlisib, Alpelisib, Ruxolitinib, Ribociclib, Erlotinib, Gemcitabine, Paclitaxel, Tamoxifen, and Trametinib. The resulting benchmark comprised 791 treatment episodes from 177 PDX models, including 343 responders and 448 non-responders, corresponding to an overall positive-response prevalence of \(0.434\). Because individual PDX models could contribute multiple treatment episodes, all episodes from a given model were assigned to the same outer fold to prevent model overlap between training and test sets. Details of the model-disjoint evaluation protocol and target-domain tuning procedure are provided in Appendix~\ref{app:pdx_protocol} and Appendix~\ref{app:target_tuning}.

\paragraph{Tumor-type composition.}
The final 10-drug benchmark included breast cancer (BRCA), cutaneous melanoma (CM), colorectal cancer (CRC), non-small-cell lung cancer (NSCLC), and pancreatic ductal adenocarcinoma (PDAC). Drug coverage was not uniform across tumor types, and some tumor--drug combinations were sparsely represented or absent. We therefore report tumor-stratified response prevalence only as a descriptive cohort characteristic rather than as an estimate of tumor-type-specific treatment efficacy. Table~\ref{tab:pdx_tumor_composition} summarizes the tumor-level cohort composition, while Table~\ref{tab:pdx_tumor_drug_composition} shows the distribution of treatment episodes across tumor--drug combinations.

\begin{table}[!htbp]
\centering
\caption{
Tumor-type composition of the shared 10-drug PDX benchmark.
Parenthetical labels denote the corresponding TCGA project codes.
Response prevalence denotes the proportion of episodes labeled CR, PR, or SD.
}
\label{tab:pdx_tumor_composition}
\small
\setlength{\tabcolsep}{6pt}
\begin{tabular}{lrrrr}
\toprule
\textbf{Tumor type}
& \textbf{Models}
& \textbf{Episodes}
& \textbf{Drugs}
& \textbf{Response prevalence} \\
\midrule
BRCA              & 38 & 227 & 6 & 0.401 \\
CM (SKCM)         & 32 & 64  & 2 & 0.375 \\
CRC (COAD/READ)   & 43 & 168 & 7 & 0.470 \\
NSCLC (LUAD/LUSC) & 25 & 124 & 5 & 0.387 \\
PDAC (PAAD)       & 39 & 208 & 6 & 0.486 \\
\midrule
Total & 177 & 791 & 10 & 0.434 \\
\bottomrule
\end{tabular}
\end{table}

\begin{table}[!htbp]
\centering
\caption{
Tumor-drug composition of the shared 10-drug PDX benchmark.
Entries denote treatment-episode counts for each tumor type-drug combination.
}
\label{tab:pdx_tumor_drug_composition}
\small
\setlength{\tabcolsep}{4pt}
\resizebox{\textwidth}{!}{
\begin{tabular}{lrrrrrrrrrrr}
\toprule
\textbf{Tumor type}
& \textbf{Alpelisib}
& \textbf{Buparlisib}
& \textbf{Erlotinib}
& \textbf{Fluorouracil}
& \textbf{Gemcitabine}
& \textbf{Paclitaxel}
& \textbf{Ribociclib}
& \textbf{Ruxolitinib}
& \textbf{Tamoxifen}
& \textbf{Trametinib}
& \textbf{Total} \\
\midrule
BRCA  & 38 & 38 & 0  & 0  & 0  & 38 & 38 & 37 & 38 & 0  & 227 \\
PDAC  & 36 & 35 & 0  & 0  & 34 & 0  & 35 & 33 & 0  & 35 & 208 \\
CRC   & 42 & 40 & 0  & 41 & 1  & 0  & 42 & 1  & 0  & 1  & 168 \\
NSCLC & 25 & 25 & 25 & 0  & 0  & 24 & 25 & 0  & 0  & 0  & 124 \\
CM    & 0  & 32 & 0  & 0  & 0  & 0  & 32 & 0  & 0  & 0  & 64  \\
\midrule
\textbf{Total}
& \textbf{141}
& \textbf{170}
& \textbf{25}
& \textbf{41}
& \textbf{35}
& \textbf{62}
& \textbf{172}
& \textbf{71}
& \textbf{38}
& \textbf{36}
& \textbf{791} \\
\bottomrule
\end{tabular}
}
\end{table}

\FloatBarrier

\subsection{Cross-benchmark Drug Coverage in Tahoe-100M}
\label{app:drug_coverage}

\paragraph{Cross-benchmark molecular coverage.}
To characterize how directly the target drugs are supported by the Tahoe-100M source chemical space, we examined all unique drugs appearing across TCGA-186, TCGA-508, and the shared PDX benchmark.
Table~\ref{tab:target_drug_chemical_coverage} reports exact Tahoe-100M support and the nearest non-identical Tahoe compound under both the frozen ChemBERTa representation used by \mname and Morgan ECFP4 fingerprint similarity.
Morgan ECFP4 fingerprints were computed using RDKit~\citep{rdkit}.

\begin{table*}[!t]
\centering
\caption{
Chemical coverage of target drugs in Tahoe-100M. Exact support indicates whether the target drug is present in Tahoe-100M by normalized drug name or canonicalized molecular structure. For each target drug, the nearest non-identical Tahoe compound is reported under ChemBERTa embedding cosine similarity and Morgan ECFP4 Tanimoto similarity. Exact target matches were excluded from nearest-neighbor searches. The two similarity measures are reported independently.
}
\label{tab:target_drug_chemical_coverage}

\resizebox{\textwidth}{!}{%
\begin{tabular}{llllr}
\toprule
\textbf{Drug} &
\textbf{Primary target / mechanism} &
\textbf{Exact Tahoe match} &
\textbf{Nearest Tahoe compound} &
\textbf{ChemBERTa cosine} \\
\midrule

Alpelisib
& PI3K$\alpha$
& Yes
& Sonidegib
& 0.906 \\

Buparlisib
& Pan-PI3K
& No
& Bimiralisib
& 0.991 \\

Cisplatin
& DNA cross-linking
& No
& Sodium Salicylate
& 0.753 \\

Docetaxel
& Microtubule stabilization
& Yes
& Docetaxel (Trihydrate)
& 0.996 \\

Erlotinib
& EGFR
& Yes
& XRK3F2
& 0.823 \\

Fluorouracil
& TYMS / DNA--RNA incorporation
& Yes
& Methylthiouracil
& 0.882 \\

Gemcitabine
& DNA incorporation / RNR
& Yes
& Decitabine
& 0.922 \\

Paclitaxel
& Microtubule stabilization
& Yes
& Vincristine
& 0.947 \\

Ribociclib
& CDK4/6
& Yes
& Palbociclib
& 0.902 \\

Ruxolitinib
& JAK1/2
& No
& Pimitespib
& 0.824 \\

Sorafenib
& RAF / VEGFR / PDGFR
& No
& Regorafenib
& 0.988 \\

Tamoxifen
& Estrogen receptor (SERM)
& No
& Diphenhydramine
& 0.837 \\

Temozolomide
& DNA methylation
& No
& Pentoxifylline
& 0.839 \\

Trametinib
& MEK1/2
& Yes
& Relugolix
& 0.938 \\

\bottomrule
\end{tabular}%
}
\vspace{0.5em}

{\footnotesize
\raggedright
\textit{Note:}
Primary target/mechanism annotations were curated from DrugBank~\citep{knox2024drugbank} and denote the principal molecular target or pharmacologic mechanism of each drug.
Similarity estimates for cisplatin should be interpreted cautiously because conventional SMILES-based representations and Morgan fingerprints are not optimized for metal complexes.
\par
}

\end{table*}

\section{Training, Implementation, and Evaluation Protocols}
\label{app:training_and_evaluation}

Unless otherwise stated, all target-domain experiments use frozen predictor transfer: the source-trained intervention encoder and transition predictor remain fixed, and only the target-domain response head is trained using response labels.

\subsection{Source-Stage Training Procedure}\label{app:training_procedure}

The~\mname pipeline consists of two stages: source-stage perturbation learning and target-stage patient response transfer.

\paragraph{Stage 1: source-stage perturbation learning.}
We first construct valid matched population pairs from the single-cell perturbation atlas. For each treated condition $(\ell,b,d,q)$, treated cells are matched to DMSO control cells from the same cell line and plate context $(\ell,b)$. Conditions are retained only if both the control and treated pools satisfy minimum cell-count requirements. During training, we sample mini-populations
\[
X \subset X_{\ell,b}^{\mathrm{DMSO}},
\qquad
Y \subset Y_{\ell,b,d,q},
\]
embed cells with the fixed pretrained encoder $E$, and train the intervention encoder $g$ and transition predictor $P_\theta$ using the distributional and delta-aware objective in Equation~\ref{eq:source_loss}. This stage learns drug- and dose-conditioned predictors that transport control-cell latent distributions toward matched treated-cell latent distributions.

\paragraph{Stage 2: target-domain response transfer.}
Pretreatment target profiles are embedded using the frozen expression pipeline, and the source-trained intervention encoder and transition predictor are applied to generate treatment-conditioned transition features. Only the response head \(h_\psi\) is trained using target-domain response labels and the binary cross-entropy loss.

This two-stage design decouples intervention learning from patient response supervision. Source-stage training uses large-scale perturbation data to learn how drugs move biological states, while patient-stage training evaluates whether these learned intervention predictors provide a useful signal when only pretreatment patient profiles and response labels are available.

\begin{algorithm}[t]
\caption{\mname source-stage training}
\label{alg:source_training}
\begin{algorithmic}[1]
\Require Valid matched populations $\mathcal{P}=\{(X_{\ell,b}^{\mathrm{DMSO}},Y_{\ell,b,d,q},u)\}$
\Require Frozen encoder $E$, intervention encoder $g$, transition predictor $P_\theta$
\For{each training step}
    \State Sample $(X_{\ell,b}^{\mathrm{DMSO}},Y_{\ell,b,d,q},u) \sim \mathcal{P}$
    \State Sample mini-populations $X \subset X_{\ell,b}^{\mathrm{DMSO}}$ and $Y \subset Y_{\ell,b,d,q}$
    \State Compute $Z_X=E(X)$ and $Z_Y=E(Y)$
    \State Encode intervention \(\alpha_u = g([m_d;\log q])\)
    \State Predict $\hat{\Delta}_i=P_\theta(z_i^X,\alpha_u)$ for each $z_i^X \in Z_X$
    \State Form $\hat{Z}_Y=\{z_i^X+\hat{\Delta}_i: z_i^X\in Z_X\}$
    \State Compute $\mathcal{L}_{\mathrm{src}}$ using Equation~\ref{eq:source_loss}
    \State Update trainable parameters using $\mathcal{L}_{\mathrm{src}}$
\EndFor
\end{algorithmic}
\end{algorithm}

\subsection{Source-Stage Objective Details}
\label{app:source_objective_details}

The source-stage objective is
\begin{equation}
\mathcal{L}_{\mathrm{src}}
=
\mathcal{L}_{\mathrm{dist}}
+
\lambda_{\mathrm{cos}}\mathcal{L}_{\mathrm{cos}}
+
\lambda_{\mathrm{mse}}\mathcal{L}_{\mathrm{mse}}
+
\lambda_{\mathrm{norm}}\mathcal{L}_{\mathrm{norm}} .
\end{equation}

\paragraph{Distributional alignment.}
We instantiate the distributional term as the biased empirical
multi-kernel maximum mean discrepancy:
\begin{equation}
\mathcal{L}_{\mathrm{dist}}
=
\mathrm{MMD}^2(\hat{Z}_Y,Z_Y).
\end{equation}
For \(A=\{a_i\}_{i=1}^{n}\) and \(B=\{b_j\}_{j=1}^{m}\),
\begin{equation}
\mathrm{MMD}^2(A,B)
=
\frac{1}{n^2}
\sum_{i,i'} k(a_i,a_{i'})
+
\frac{1}{m^2}
\sum_{j,j'} k(b_j,b_{j'})
-
\frac{2}{nm}
\sum_{i,j} k(a_i,b_j).
\end{equation}

We use a multi-kernel Gaussian kernel,
\begin{equation}
k(x,x')
=
\sum_{\sigma\in\mathcal{S}}
\exp\left(
-\frac{\|x-x'\|_2^2}{2\sigma^2}
\right),
\end{equation}
with bandwidth set
\[
\mathcal{S}=\{8,16,32,64,128\}.
\]

\paragraph{Transition-direction alignment.}
The cosine term aligns the direction of the predicted and empirical
population transitions:
\begin{equation}
\mathcal{L}_{\mathrm{cos}}
=
1-
\frac{
\hat{\Delta}_{c,u}^{\top}\Delta_{c,u}
}{
\left(\|\hat{\Delta}_{c,u}\|_2+\epsilon\right)
\left(\|\Delta_{c,u}\|_2+\epsilon\right)
},
\end{equation}
where \(\epsilon>0\) is a small numerical constant used to avoid division by zero.

\paragraph{Transition-displacement alignment.}
The vector-displacement term is
\begin{equation}
\mathcal{L}_{\mathrm{mse}}
=
\frac{1}{d}
\left\|
\hat{\Delta}_{c,u}
-
\Delta_{c,u}
\right\|_2^2.
\end{equation}

\paragraph{Transition-norm alignment.}
We additionally constrain the magnitude of the predicted population
transition using
\begin{equation}
\mathcal{L}_{\mathrm{norm}}
=
\left(
\left\|\hat{\Delta}_{c,u}\right\|_2
-
\left\|\Delta_{c,u}\right\|_2
\right)^2 .
\end{equation}
This term penalizes discrepancies in transition magnitude independently
of transition direction.

These terms discourage generic control-to-treated shifts and promote intervention-specific transition predictors.

\paragraph{Loss-weight selection.}
We use \(\lambda_{\mathrm{cos}}=0.1\), \(\lambda_{\mathrm{mse}}=3.0\), and \(\lambda_{\mathrm{norm}}=10^{-3}\). The weights were selected using a source-stage validation grid without access to target-domain response labels. The loss-weight search space and selection procedure are reported in Appendix~\ref{app:source_loss_selection}.

\subsection{Source-Stage Validation Protocol}
\label{app:source_validation_protocol}

We evaluate source-stage generalization using two complementary validation settings constructed from the valid Tahoe-100M source rows. Each source row represents a matched population pair indexed by cell line, plate, drug, and dose. We first construct the held-out-drug split and then construct the random held-out split from the remaining rows.

\paragraph{Held-out-drug validation.}
Eligible drugs are randomly ordered using split seed~0. Complete drugs are then withheld, together with all of their associated cell-line, plate, and dose rows, until the cumulative number of withheld rows reaches at least 10\% of the valid source rows. Because drugs are removed as indivisible groups, the resulting fraction can slightly exceed 10\%. The two drugs in TCGA-186 with exact Tahoe-100M support, 5-Fluorouracil and Gemcitabine, are protected from held-out-drug selection so that their source perturbation observations remain available for the patient-transfer analysis. This procedure yields 7,488 held-out rows spanning 42 drugs, corresponding to 10.07\% of the 74,396 valid source rows.

This split evaluates \emph{zero-shot drug generalization in the source perturbation domain}: the model receives the held-out compounds' molecular representations and doses at evaluation time, but no matched perturbation populations for those compounds are used during source training. It therefore tests whether the learned drug-conditioned transition operator transfers to compounds without compound-specific perturbation supervision. It does not test the absence of all information about the compound, because the frozen ChemBERTa representation remains available.

\paragraph{Random held-out validation.}
After removing the held-out-drug rows, 10\% of the remaining rows are sampled uniformly at random using random-validation seed~1. This produces 6,691 random held-out rows; the remaining 60,217 rows are used for source training. The training and random held-out partitions both contain 336 drugs, so this split is not drug-disjoint.

The random held-out split evaluates generalization to unseen \emph{context-intervention rows} for drugs represented during training. Because rows are sampled independently rather than grouped by cell line, plate, drug, or dose, a held-out row may share its drug, dose, cell line, or plate with training rows. Accordingly, this setting primarily measures within-support interpolation across observed perturbation factors and should not be interpreted as a strict generalization to unseen drugs, unseen cell lines, unseen plates, or unseen doses.

\paragraph{Interpretation of the comparison.}
The two splits distinguish condition-level generalization from compound-level generalization. Strong performance on random held-out rows indicates that the model can recover transitions for new combinations drawn from the observed source support. Performance on held-out drugs is a stricter test of whether molecular drug representations support transfer to compounds with no source-stage perturbation observations. The substantially lower transition cosine and higher MMD and delta MSE on held-out drugs therefore quantify the limitation of zero-shot cross-drug transfer rather than a failure to fit seen-drug perturbation conditions.

Source-stage performance is evaluated using MMD between predicted and observed treated populations, cosine similarity between predicted and empirical population transitions, and delta MSE. The split sizes are summarized below.


\subsection{Model Architecture and Hyperparameters}
\label{app:architecture_hyperparameters}

We use a frozen scFoundation encoder for expression profiles. Cells are embedded with a frozen scFoundation encoder, and the concatenated cell embedding has dimension $d=3072$ in our implementation. Drug features are obtained from frozen ChemBERTa embeddings (i.e., ``seyonec/ChemBERTa-zinc-base-v1'') and combined with a log-dose feature before the trainable intervention encoder $g$ produces the intervention embedding $\alpha_u$.

Table~\ref{tab:architecture_hyperparameters} summarizes the architecture and training hyperparameters used in the main~\mname experiments. Source-stage parameters are selected exclusively using Tahoe-100M validation data, whereas target-domain response-head and representation settings are described separately in Appendix~\ref{app:target_tuning}.

\begin{table}[t]
\centering
\caption{
Architecture and hyperparameter details for~\mname.
}
\label{tab:architecture_hyperparameters}

\resizebox{\columnwidth}{!}{
\begin{tabular}{ll}
\toprule
\textbf{Component} & \textbf{Configuration} \\
\midrule

\multicolumn{2}{l}{\textbf{Source-stage transition learning}} \\
\midrule

Cell encoder
& Frozen scFoundation, \(d=3072\) \\

Drug encoder
& Frozen ChemBERTa, \(d=768\) \\

Intervention encoder \(g\)
& Input \(768+1\); hidden/output dim \(512/512\) \\

Transition predictor \(P_\theta\)
& Input \(3072+512\), hidden dim \(1024\), output dim \(3072\) \\

Dropout
& \(0.1\) \\

Optimizer
& AdamW, learning rate \(10^{-4}\), weight decay \(10^{-4}\) \\

Batch construction
& \(32\) conditions; \(128\) cells per population \\

Training
& \(100\) epochs \\

Loss weights
& \(\lambda_{\cos}=0.1,\;
   \lambda_{\mathrm{mse}}=3.0,\;
   \lambda_{\mathrm{norm}}=10^{-3}\) \\

\midrule
\multicolumn{2}{l}{\textbf{Target-domain response prediction}} \\
\midrule

Frozen components
& scFoundation, ChemBERTa, \(g\), and \(P_\theta\) \\

Response input
& \([z;m_d;\hat{\Delta}_u]\) \\

Response head \(h_\psi\)
& LayerNorm-Linear-GELU-Dropout-Linear \\

Training
& AdamW with class-weighted binary cross-entropy \\

\midrule
\multicolumn{2}{l}{\textit{TCGA-186}} \\
\midrule

Hidden dim / dropout
& \(128 / 0.7\) \\

Learning rate / weight decay
& \(10^{-3} / 10^{-3}\) \\

Inference dose
& \(0.05\,\mu\mathrm{M}\) \\

\midrule
\multicolumn{2}{l}{\textit{TCGA-508}} \\
\midrule

Hidden dim / dropout
& \(64 / 0.4\) \\

Learning rate / weight decay
& \(3\times10^{-4} / 0.1\) \\

Inference dose
& \(0.5\,\mu\mathrm{M}\) \\

\midrule
\multicolumn{2}{l}{\textit{PDX shared 10-drug benchmark}} \\
\midrule

Hidden dim / dropout
& \(128 / 0.2\) \\

Learning rate / weight decay
& \(10^{-3} / 10^{-4}\) \\

Inference dose
& \(0.05\,\mu\mathrm{M}\) \\

\bottomrule
\end{tabular}
}
\end{table}

\subsection{Baseline Implementations}
\label{app:baseline_definitions}

We compare \mname against four representative baselines for patient drug-response prediction and cell-line-to-patient transfer: CODE-AE~\citep{he2022context}, TransDRP~\citep{liu2025knowledge}, DeepSADR~\citep{zhang2026deepsadr}, and WISER~\citep{shubham2024wiser}. These methods span domain adaptation, transfer learning, and weak supervision for leveraging pharmacogenomic cell-line data in patient response prediction.

The baselines were originally developed for transferring information from cell-line drug-response screens and do not natively use large single-cell perturbation atlases such as Tahoe-100M. We discuss this difference in available pretraining data, together with the corresponding controls, in Section~\ref{sec:patient_response_prediction} and Table~\ref{tab:tcga_ablation}.

\paragraph{CODE-AE.~\citep{he2022context}}
CODE-AE learns shared and domain-specific gene-expression representations and aligns the shared latent space between source cell lines and target patients using domain-adversarial training. A drug-response predictor is subsequently trained in the learned representation space. We use the released implementation and adapt the expression alignment, sample indexing, and evaluation interface to our benchmark.

\paragraph{TransDRP.~\citep{liu2025knowledge}}
TransDRP transfers cell-line drug-response predictors to patient samples through drug-aware response modeling and global-local domain-adversarial alignment. We retain the original training and adaptation procedure while harmonizing gene features, drug identifiers, evaluation splits, and prediction outputs for the common benchmark.

\paragraph{DeepSADR.~\citep{zhang2026deepsadr}}
DeepSADR transfers a response model pretrained on lower-fidelity cell-line data to higher-fidelity patient response prediction through supervised fine-tuning. We use the released pretrained model and fine-tune it only on samples in the non-held-out training partition. Model selection is performed without using response labels from the held-out evaluation partition.

\paragraph{WISER.~\citep{shubham2024wiser}}
WISER combines domain-invariant representation learning with weak supervision. It generates candidate predictions for target-domain samples, selects reliable pseudo-labeled examples, and combines them with labeled source-domain data to train the final response predictor. We preserve the released weak-supervision pipeline while enforcing the common evaluation splits.

\paragraph{Evaluation protocol.}
Whenever available, we use implementations released by the original authors and preserve each method's original architecture, objectives, and method-specific supervision regime. Benchmark-specific modifications are limited to data formatting, gene and drug identifier harmonization, split handling, and prediction export. 

For TCGA-186, we retain the baseline configurations used in the corresponding released implementations and do not tune them using held-out patient response labels. For TCGA-508 and PDX, each method is retrained or adapted to the corresponding target cohort according to its native training procedure. When model selection, early stopping, or target-domain tuning is required, it is performed using only the non-held-out portion of the corresponding evaluation split. Held-out response labels are not used for training, adaptation, pseudo-labeling, or model selection. 

We do not require identical hyperparameters across methods; instead, all methods share the same held-out evaluation partitions and test-data exclusion boundary while retaining their method-specific training procedures. TCGA-186 results are summarized across five patient-disjoint splits, TCGA-508 results across five patient-grouped folds, and PDX results across five model-disjoint outer folds.

\subsection{TCGA-186 Seen-Drug, Held-Out-Patient Evaluation}
\label{app:seen_drug_protocol}

For TCGA-186, we evaluate response prediction in a seen-drug, held-out-patient setting.
The benchmark contains 186 patient-drug response records from 186 unique patients, and the five evaluated drugs are represented in the patient-stage training data.
Thus, the task measures generalization to previously unseen patients for drugs observed during patient-stage training, rather than generalization to entirely unseen compounds.

We construct five patient-disjoint train/test splits using different random seeds and use the same splits for all methods.
Because each patient contributes a single response record in TCGA-186, record-level and patient-level separation are equivalent for this cohort.
For every split, patients assigned to the test partition are excluded from patient-stage model fitting and model selection.

Performance is computed on the held-out patients for each split.
We report both aggregate performance and drug-stratified performance, with mean and standard deviation summarized across the five random seeds.
For drug-stratified analyses, AUROC and AUPRC are computed independently within each treatment condition.

\subsection{PDX Model-Disjoint Evaluation}
\label{app:pdx_protocol}

We evaluate the PDX cohort using five outer folds grouped by PDX model identifier.
All treatment episodes from a given PDX model are assigned to the same fold, ensuring that no PDX model contributes observations to both the training and test partitions of a given evaluation.
The single PDX model lacking a tumor-type annotation is excluded from the benchmark.

The primary cross-method comparison is performed on the 10 drugs shared across the evaluated methods:
Fluorouracil, Buparlisib, Alpelisib, Ruxolitinib, Ribociclib, Erlotinib, Gemcitabine, Paclitaxel, Tamoxifen, and Trametinib.
This shared subset contains 791 model-drug treatment episodes from 177 PDX models.

For each outer fold, the held-out PDX models are excluded from response-model fitting.
Any method-specific target-domain adaptation, pseudo-labeling, feature normalization, or supervised patient/PDX fine-tuning is restricted to the non-held-out models according to the native training procedure of each method.
Thus, the methods retain their original supervision mechanisms while sharing the same model-disjoint evaluation boundary.

For \mname, the response predictor uses the concatenation of the pretreatment PDX embedding, ChemBERTa drug embedding, and frozen predicted perturbation transition.
Transition features are evaluated at inference doses
\(q \in \{0.05, 0.5, 5.0\}\,\mu\mathrm{M}\).
The pretrained perturbation model remains frozen during PDX response-model training.

Drug-specific AUROC and AUPRC are computed separately within each held-out outer fold and summarized as mean $\pm$ standard deviation across the five folds.
The drug-stratified comparison therefore reflects variability across model-disjoint PDX test partitions rather than variability across random initialization seeds.
Aggregate and drug-specific results are reported separately.

\section{Additional Results and Diagnostics}
\label{app:additional_results}

\subsection{Source-Target Embedding Distribution Diagnostics}
\label{app:source_target_embedding_diagnostics}

Although Tahoe-100M single-cell profiles and TCGA bulk RNA-seq profiles are processed using the same frozen scFoundation checkpoint and embedding pipeline, this does not imply that their latent distributions are aligned. We therefore compared TCGA patient embeddings with Tahoe DMSO cell embeddings using balanced joint principal component analysis.

PCA was fit using all 186 TCGA patient embeddings and an equal-sized random sample of 186 Tahoe DMSO cells selected with a fixed random seed, preventing the substantially larger Tahoe population from dominating the fitted components. The remaining sampled Tahoe cells and cell-line centroids were then projected into the fitted PCA space. We repeated the analysis using raw embeddings, sample-wise \(L_2\)-normalized embeddings, and feature-wise standardized embeddings.

\begin{figure*}[t]
    \centering
    \includegraphics[width=\textwidth]
    {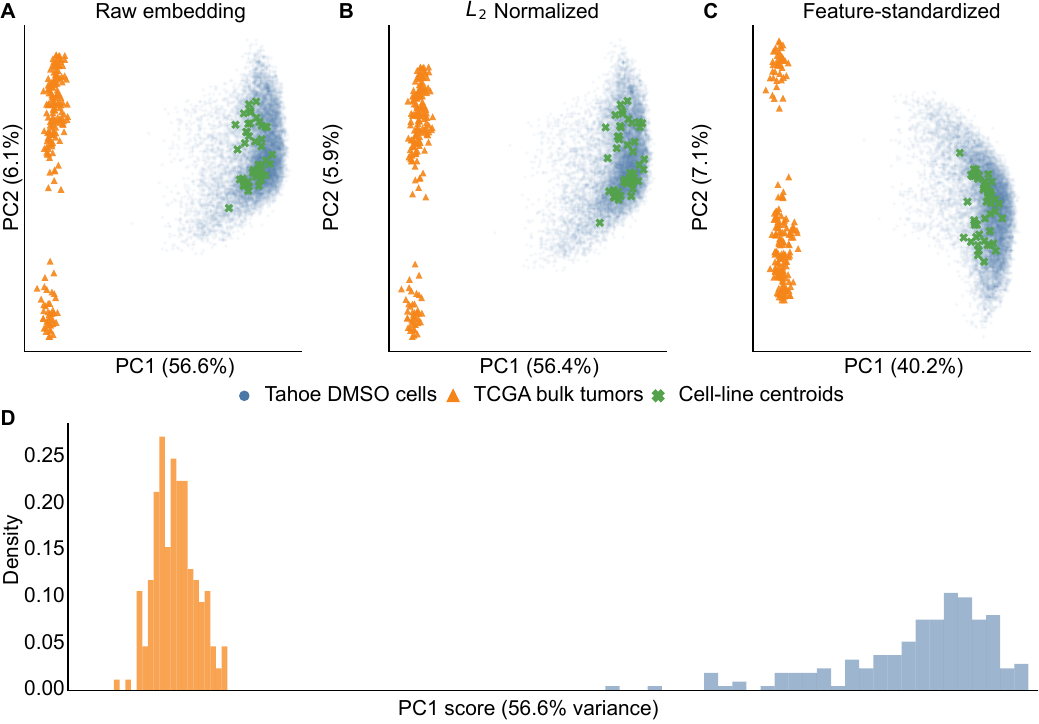}
    \caption{
    \textbf{Source-target embedding distribution diagnostics.}
    \textbf{A-C.}
    Balanced joint PCA of TCGA bulk-tumor embeddings, Tahoe DMSO
    single-cell embeddings, and Tahoe cell-line centroids using
    \textbf{A} raw embeddings,
    \textbf{B} sample-wise \(L_2\)-normalized embeddings, and
    \textbf{C} feature-wise standardized embeddings.
    \textbf{D.}
    Distribution of raw-embedding PC1 scores for TCGA patients and
    Tahoe DMSO cells.
    PCA was fit using equal numbers of TCGA patients and Tahoe cells.
    }
    \label{fig:source_target_embedding_shift}
\end{figure*}

As shown in Figure~\ref{fig:source_target_embedding_shift}, the TCGA and Tahoe embeddings remain clearly separated under all three representation variants, indicating that the source-target difference is not explained solely by embedding norm or individual latent-coordinate scale. In the raw-embedding analysis, PC1 explained 56.6\% of the variance and captured the dominant source-target mean offset. A five-fold cross-validated logistic-regression domain classifier, fit on the balanced TCGA-Tahoe sample, distinguished the two domains with an AUROC and AUPRC of 1.0. At the same time, a balanced 20-nearest-neighbor analysis yielded mean cross-domain neighbor fractions of 0.000 for TCGA patients and 0.00027 for Tahoe cells.

These diagnostics reveal a substantial shift in the bulk-single-cell distribution in the frozen scFoundation output space. We therefore interpret the patient-stage experiments as tests of functional transfer under domain shift, rather than as evidence that patient tumors and source cell lines occupy an aligned or domain-invariant static embedding space.

\subsection{Dose Sensitivity Across Target Cohorts}
\label{app:dose_sensitivity}

Because clinical and in vivo treatment exposures cannot be directly mapped to the Tahoe-100M micromolar dose scale, we treat \(q \in \{0.05, 0.5, 5.0\}\,\mu\mathrm{M}\) as alternative intervention-representation settings rather than calibrated therapeutic exposures. Unless otherwise stated, the frozen predictor, cohort, evaluation splits, feature-processing pipeline, response-model architecture, and training procedure were held fixed across doses.

\paragraph{TCGA-186.}
Clinical treatment records in TCGA-186 do not provide standardized dose
values that can be mapped to the Tahoe-100M dose scale. We therefore treat
the inference dose as an intervention-representation setting rather than a
calibrated clinical exposure. Using the fixed final response-model
configuration, we evaluated the three Tahoe-100M dose settings
\(q\in\{0.05,0.5,5.0\}\,\mu\mathrm{M}\).
The \(0.05\,\mu\mathrm{M}\) setting achieved the highest mean AUROC and
AUPRC for each of the five TCGA-186 drugs and is therefore used for the
primary TCGA-186 results.

\begin{table}[!htbp]
\centering
\caption{
Drug-wise dose sensitivity of \mname on TCGA-186 under the fixed final
response-model configuration. Values are mean \(\pm\) standard deviation
across five held-out-patient evaluations. Bold indicates the highest mean
and underline the second-highest mean within each drug and metric.
The inference dose is treated as a representation setting rather than a
calibrated clinical exposure.
}
\label{tab:patient_dose_sensitivity}

\small
\setlength{\tabcolsep}{3.2pt}
\renewcommand{\arraystretch}{1.00}

\begin{tabular}{lcccccc}
\toprule
& \multicolumn{3}{c}{\textbf{AUROC}}
& \multicolumn{3}{c}{\textbf{AUPRC}} \\
\cmidrule(lr){2-4}
\cmidrule(lr){5-7}

\textbf{Drug}
& \textbf{0.05}
& \textbf{0.5}
& \textbf{5.0}
& \textbf{0.05}
& \textbf{0.5}
& \textbf{5.0} \\
\midrule

Cisplatin
& \textbf{\meanstdsmall{0.684}{0.056}}
& \meanstdsmall{0.641}{0.086}
& \underline{\meanstdsmall{0.647}{0.021}}
& \textbf{\meanstdsmall{0.733}{0.093}}
& \meanstdsmall{0.705}{0.111}
& \underline{\meanstdsmall{0.711}{0.046}} \\

Fluorouracil
& \textbf{\meanstdsmall{0.674}{0.104}}
& \meanstdsmall{0.560}{0.214}
& \underline{\meanstdsmall{0.600}{0.197}}
& \textbf{\meanstdsmall{0.645}{0.134}}
& \meanstdsmall{0.526}{0.173}
& \underline{\meanstdsmall{0.576}{0.197}} \\

Gemcitabine
& \textbf{\meanstdsmall{0.551}{0.082}}
& \underline{\meanstdsmall{0.507}{0.100}}
& \meanstdsmall{0.491}{0.124}
& \textbf{\meanstdsmall{0.530}{0.055}}
& \meanstdsmall{0.511}{0.081}
& \underline{\meanstdsmall{0.522}{0.106}} \\

Sorafenib
& \textbf{\meanstdsmall{0.753}{0.213}}
& \meanstdsmall{0.667}{0.162}
& \underline{\meanstdsmall{0.740}{0.201}}
& \textbf{\meanstdsmall{0.801}{0.174}}
& \meanstdsmall{0.704}{0.160}
& \underline{\meanstdsmall{0.773}{0.172}} \\

Temozolomide
& \textbf{\meanstdsmall{0.653}{0.142}}
& \meanstdsmall{0.638}{0.133}
& \underline{\meanstdsmall{0.647}{0.164}}
& \textbf{\meanstdsmall{0.680}{0.107}}
& \underline{\meanstdsmall{0.655}{0.086}}
& \meanstdsmall{0.651}{0.115} \\

\bottomrule
\end{tabular}
\end{table}

\subsection{Cross-cohort heterogeneity in transition-magnitude associations}
\label{app:transition_magnitude}

To assess whether the response-associated transition geometry observed
in TCGA-508 generalized across target cohorts, we performed an
exploratory drug-stratified analysis of predicted transition magnitude
in TCGA-508, TCGA-186, and PDX.
For each cohort--drug stratum, we computed the standardized difference
in transition magnitude between responders and non-responders, with
negative values indicating smaller predicted transitions in responders.

\begin{figure}[t]
    \centering
    \includegraphics[width=0.78\linewidth]
    {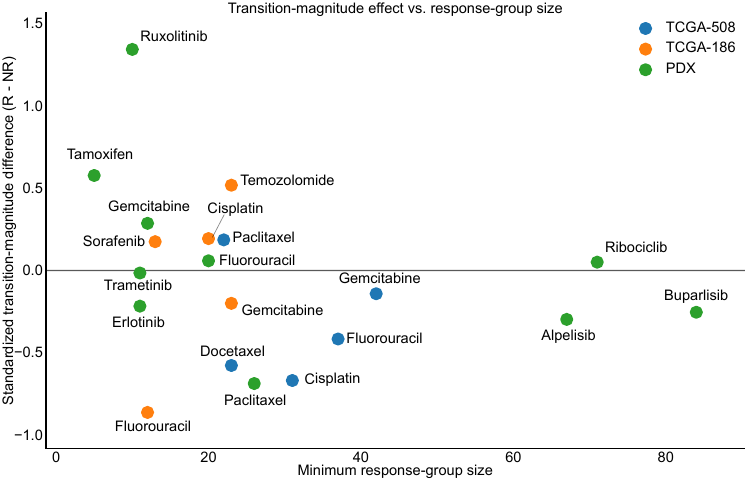}
    \caption{
    \textbf{Transition-magnitude effect versus response-group size.}
    Each point represents one cohort--drug stratum and shows the
    standardized difference in predicted transition magnitude between
    responders and non-responders.
    Negative values indicate smaller predicted transitions in responders.
    Effect estimates were more variable in sparsely represented strata,
    whereas larger strata tended to show negative or near-zero effects.
    This analysis is descriptive and is not used as primary evidence for
    a universal magnitude--response association.
    }
    \label{fig:transition_magnitude_group_size}
\end{figure}

Across all 20 primary-dose cohort--drug strata, the random-effects
pooled standardized effect was small and not statistically resolved
(\(g=-0.045\), 95\% CI \([-0.269,0.180]\)),
with substantial between-stratum heterogeneity
(\(I^2=63.0\%\)).
The descriptive pattern in
Figure~\ref{fig:transition_magnitude_group_size} shows that the most
extreme positive estimates occurred primarily in smaller response
strata, whereas several of the better-represented strata showed
negative or near-zero effects.

In an exploratory sample-size sensitivity analysis, restricting to
strata with at least 25 observations in both response groups yielded a
negative pooled effect (\(g=-0.270\), 95\% CI
\([-0.474,-0.067]\)); a threshold of 30 observations yielded a similar
estimate (\(g=-0.283\), 95\% CI \([-0.522,-0.043]\)).
Because these thresholds were examined post hoc, these results are
treated as sensitivity analyses rather than evidence for a universal
response-associated transition signature.

\section{Source-Stage Model Selection and Hyperparameter Tuning}
\label{app:source_loss_selection}

All source-stage model selection was performed using Tahoe-100M source validation data only; no TCGA or PDX response labels were used for selecting the source objective, molecular representation, or loss weights.

\paragraph{Validation metrics.}
For each validation condition, we compare the predicted population-level transition,
\[
\hat{\Delta}
=
\frac{1}{N}\sum_i \hat{z}^{Y}_i
-
\frac{1}{N}\sum_i z^{X}_i,
\]
with the empirical population-level transition,
\[
\Delta
=
\frac{1}{N}\sum_i z^{Y}_i
-
\frac{1}{N}\sum_i z^{X}_i.
\]
Our primary source-stage selection metric is the cosine similarity
\[
s_{\mathrm{cos}}
=
\cos(\hat{\Delta},\Delta).
\]
Higher values indicate better agreement between the predicted and empirical transition directions. During training, the corresponding cosine loss is defined as
\[
\mathcal{L}_{\mathrm{cos}}
=
1-s_{\mathrm{cos}},
\]
and is therefore minimized. We report cosine similarity rather than cosine loss for validation results.

For source-stage model selection, we evaluate each configuration on randomly held-out source conditions. These conditions correspond to unseen cell-line-plate-drug-dose combinations drawn from drugs represented during source training. This validation setting therefore measures generalization to unseen source conditions within the observed drug support, rather than generalization to entirely unseen compounds. Model selection is based on transition cosine similarity between the predicted and empirical population-level transitions, with higher values indicating better agreement.

\paragraph{Loss-weight sweep.}
We first evaluated the source objective
\[
\mathcal{L}
=
\lambda_{\mathrm{dist}}\mathcal{L}_{\mathrm{dist}}
+
\lambda_{\mathrm{cos}}\mathcal{L}_{\mathrm{cos}}
+
\lambda_{\mathrm{mse}}\mathcal{L}_{\mathrm{mse}}
+
\lambda_{\mathrm{norm}}\mathcal{L}_{\mathrm{norm}},
\]
with \(\lambda_{\mathrm{dist}}=1\).
For the initial MMD-based sweep, we evaluated
\[
\lambda_{\mathrm{cos}}\in\{0, 0.1, 0.3, 1.0\},
\qquad
\lambda_{\mathrm{mse}}\in\{0,0.3,1.0,3.0\},
\qquad
\lambda_{\mathrm{norm}}\in\{0, 10^{-3}\},
\]
yielding 32 configurations. Each configuration was trained for 300 optimization steps under otherwise identical source-stage settings.

\paragraph{Distributional objective and molecular representation.}
We subsequently compared maximum mean discrepancy (MMD) with entropically regularized Sinkhorn optimal transport as the population-level distributional loss. We additionally compared three pretrained molecular representations: the original ChemBERTa representation, ChemBERTa-77M-MTR, and MoLFormer. The original ChemBERTa, ChemBERTa-77M-MTR, and MoLFormer were evaluated on the same 376-drug source subset and therefore permit a matched comparison of molecular representation and distributional objective.

\begin{table}[b]
\centering
\caption{
Best source-stage configuration within each molecular representation and distributional-loss combination. Performance is measured by transition cosine similarity on randomly held-out source conditions; higher is better. 
}
\label{tab:source_model_selection}
\small
\begin{tabular}{llcccc}
\toprule
Drug representation &
Distribution loss &
$\lambda_{\cos}$ &
$\lambda_{\mathrm{mse}}$ &
$\lambda_{\mathrm{norm}}$ &
Held-out condition cosine $\uparrow$ \\
\midrule
ChemBERTa-zinc-base-v1 & MMD
& 0.1 & 3.0 & $10^{-3}$
& \textbf{0.2099} \\

ChemBERTa-zinc-base-v1 & Sinkhorn
& 0.3 & 1.0 & 0
& 0.1906 \\

ChemBERTa-77M-MTR & MMD
& 1.0 & 1.0 & $10^{-3}$
& 0.2074 \\

ChemBERTa-77M-MTR & Sinkhorn
& 0.3 & 3.0 & $10^{-3}$
& 0.1989 \\

MoLFormer-XL & MMD
& 1.0 & 1.0 & $10^{-3}$
& 0.1856 \\

MoLFormer-XL & Sinkhorn
& 1.0 & 3.0 & $10^{-3}$
& 0.1938 \\
\bottomrule
\end{tabular}
\end{table}

\paragraph{Selection criterion.}
We select source-stage configurations using transition cosine similarity on randomly held-out source conditions. Higher cosine similarity indicates better agreement between the predicted and empirical population-level transition. No target-domain response labels or target-domain performance metrics were used for source-stage model selection. Table~\ref{tab:source_model_selection} summarizes the best configuration within each molecular-representation and distributional-loss combination under this criterion. Based on the random held-out condition criterion, we selected ChemBERTa-zinc-base-v1 with MMD and \(\lambda_{\cos}=0.1\), \(\lambda_{\mathrm{mse}}=3.0\), and \(\lambda_{\mathrm{norm}}=10^{-3}\) for the final source-stage model.

\section{Target-Domain Hyperparameter and Representation-Setting Selection}
\label{app:target_tuning}

Target-domain model selection was performed separately from source-stage model selection. Throughout target-domain evaluation, the pretrained intervention encoder and transition predictor remained frozen. Only the response head and target-domain representation settings were selected using data from the non-held-out portion of the corresponding target benchmark.

\paragraph{Response-head hyperparameters.}
We treated the hidden dimension, dropout rate, learning rate, and weight decay of the response head as target-domain hyperparameters. The response head takes the concatenated representation \( [z_p; m_d; \hat{\Delta}_{p,u}] \) as input and outputs the predicted treatment-response probability. Candidate configurations were evaluated using development data under the same patient-level or model-level grouping constraints used for the corresponding benchmark, without using response labels from the outer held-out evaluation partition.

For the primary TCGA-186 benchmark, the final response-head configuration used a hidden dimension of 128, dropout of 0.7, learning rate of \(10^{-3}\), and weight decay of \(10^{-3}\). Training used a maximum of 500 epochs with an early-stopping patience of 50 epochs.

For TCGA-508, the final response-head configuration used a hidden dimension of 64, dropout of 0.4, learning rate of \(3\times10^{-4}\), and weight decay of 0.1, with an inference dose of \(0.5\,\mu\mathrm{M}\).

\paragraph{Inference-dose setting.} The perturbation dose supplied to the frozen source-stage intervention encoder is treated as a representation setting rather than as a calibrated clinical or in vivo exposure. Because clinical and PDX treatment exposures cannot be directly mapped to the micromolar dose scale used in Tahoe-100M, we evaluate a small set of source-scale inference doses and report dose-sensitivity analyses separately for each target benchmark.

For TCGA-186, the primary reported results use an inference dose of \(0.05\,\mu\mathrm{M}\). This setting is used only to construct the frozen transition representation and should not be interpreted as an estimated patient-specific or clinically calibrated drug concentration.

\paragraph{Evaluation protocol.} Target-domain selection is kept separate from the source-stage model-selection procedure described in Section~\ref{app:source_loss_selection}. Source-stage objectives, molecular representations, and loss weights are selected exclusively using Tahoe-100M validation conditions, whereas target-domain response-head hyperparameters and representation settings are selected using only the corresponding target-domain development data. The outer held-out partitions are reserved for final performance estimation.

Unless otherwise stated, all reported target-domain comparisons use the same data partitions, response definitions, and evaluation metrics as the corresponding benchmark experiments. Detailed dose-sensitivity analyses for TCGA-508 and PDX are reported in Appendices~\ref{app:tcga508_dose_sensitivity} and ~\ref{app:pdx_shared_dose_sensitivity}, respectively.

\section{TCGA-508 Dose Sensitivity Under the Fixed Final Response Model}
\label{app:tcga508_dose_sensitivity}

To isolate the effect of the intervention-representation dose, we repeated the TCGA-508 evaluation while holding all other modeling choices fixed to the final setting used in the main patient-grouped benchmark. Specifically, we used the same pretrained source checkpoint, response-head architecture and hyperparameters, patient-grouped outer folds, inner patient-grouped validation procedure, feature standardization, and early-stopping procedure, and varied only the Tahoe-100M inference dose over \(q \in \{0.05, 0.5, 5.0\}\,\mu\mathrm{M}\). Thus, the comparison below measures dose sensitivity without reselecting the response-model configuration separately for each dose.

Table~\ref{tab:tcga508_dose_sensitivity} reports mean \(\pm\) standard deviation across the five patient-grouped evaluation folds. Overall performance was stable across the three settings, with \(0.5\,\mu\mathrm{M}\) yielding the highest mean AUROC and AUPRC. Drug-specific sensitivity was more heterogeneous: Cisplatin favored \(0.05\,\mu\mathrm{M}\), Fluorouracil favored \(5.0\,\mu\mathrm{M}\), and the remaining drugs generally showed their strongest or near-strongest performance at \(0.5\,\mu\mathrm{M}\). These results reinforce that the Tahoe-100M dose is best interpreted as an intervention-representation setting rather than a calibrated clinical exposure.

\begin{table*}[t]
\centering
\caption{
Drug-wise dose sensitivity of \mname on TCGA-508 under the fixed final response-model setting. Values are mean $\pm$ standard deviation across five patient-grouped evaluation folds. Bold indicates the best mean within each drug and metric across the three inference doses, underline indicates the second-best mean, and $\dagger$ indicates the third-best mean. The inference dose is treated as a representation setting and is not calibrated to clinical treatment exposure.
}
\label{tab:tcga508_dose_sensitivity}
\scriptsize
\setlength{\tabcolsep}{5.5pt}
\renewcommand{\arraystretch}{1.0}
\begin{tabular}{lcccccc}
\toprule
& \multicolumn{3}{c}{\textbf{AUROC}} & \multicolumn{3}{c}{\textbf{AUPRC }} \\
\cmidrule(lr){2-4}\cmidrule(lr){5-7}
\textbf{Drug} & \textbf{0.05} & \textbf{0.5} & \textbf{5.0} & \textbf{0.05} & \textbf{0.5} & \textbf{5.0} \\
\midrule
Overall
& \meanstdsmall{0.682}{0.054}$^{\dagger}$ & \textbf{\meanstdsmall{0.692}{0.055}} & \underline{\meanstdsmall{0.684}{0.062}}
& \meanstdsmall{0.783}{0.063}$^{\dagger}$ & \textbf{\meanstdsmall{0.787}{0.055}} & \underline{\meanstdsmall{0.785}{0.056}} \\
Cisplatin
& \textbf{\meanstdsmall{0.686}{0.166}} & \meanstdsmall{0.631}{0.168}$^{\dagger}$ & \underline{\meanstdsmall{0.638}{0.140}}
& \textbf{\meanstdsmall{0.872}{0.105}} & \meanstdsmall{0.864}{0.085}$^{\dagger}$ & \underline{\meanstdsmall{0.869}{0.075}} \\
Docetaxel
& \underline{\meanstdsmall{0.709}{0.215}} & \textbf{\meanstdsmall{0.711}{0.141}} & \meanstdsmall{0.687}{0.173}$^{\dagger}$
& \textbf{\meanstdsmall{0.816}{0.171}} & \underline{\meanstdsmall{0.802}{0.112}} & \meanstdsmall{0.772}{0.150}$^{\dagger}$ \\
Fluorouracil
& \meanstdsmall{0.552}{0.143}$^{\dagger}$ & \underline{\meanstdsmall{0.562}{0.116}} & \textbf{\meanstdsmall{0.568}{0.076}}
& \meanstdsmall{0.714}{0.120}$^{\dagger}$ & \underline{\meanstdsmall{0.748}{0.098}} & \textbf{\meanstdsmall{0.750}{0.080}} \\
Gemcitabine
& \meanstdsmall{0.455}{0.095}$^{\dagger}$ & \textbf{\meanstdsmall{0.472}{0.095}} & \underline{\meanstdsmall{0.460}{0.108}}
& \underline{\meanstdsmall{0.387}{0.053}} & \textbf{\meanstdsmall{0.395}{0.052}} & \meanstdsmall{0.385}{0.058}$^{\dagger}$ \\
Paclitaxel
& \meanstdsmall{0.631}{0.075}$^{\dagger}$ & \textbf{\meanstdsmall{0.715}{0.135}} & \underline{\meanstdsmall{0.638}{0.185}}
& \meanstdsmall{0.822}{0.086}$^{\dagger}$ & \textbf{\meanstdsmall{0.842}{0.081}} & \underline{\meanstdsmall{0.840}{0.072}} \\
\bottomrule
\end{tabular}
\end{table*}

\section{PDX Dose Sensitivity}
\label{app:pdx_shared_dose_sensitivity}

We evaluated \mname at all three Tahoe-100M inference-dose settings, \(q \in \{0.05, 0.5, 5.0\}\,\mu\mathrm{M}\), on the same shared 10-drug PDX subset used for the cross-method comparison. The inference dose is treated only as a representation setting and is not calibrated to in vivo PDX exposure. Table~\ref{tab:pdx_shared_dose_sensitivity} reports drug-specific AUROC and AUPRC  as mean $\pm$ standard deviation across the five model-disjoint outer folds. The \mname entries in the cross-method PDX results (Table~\ref{tab:pdx_drugwise}) correspond to the \(0.05\,\mu\mathrm{M}\) inference setting. Dose sensitivity was drug dependent, and no single inference dose was uniformly optimal across all ten drugs.

\begin{table*}[t]
\centering
\caption{
Drug-wise dose sensitivity of \mname on the shared 10-drug PDX benchmark. Values are mean $\pm$ standard deviation across five model-disjoint outer folds. Bold indicates the best mean within each drug and metric across the three inference doses, underline indicates the second-best mean, and $\dagger$ indicates the third-best mean. The tested doses are representation settings inherited from Tahoe-100M and should not be interpreted as calibrated PDX exposures.
}
\label{tab:pdx_shared_dose_sensitivity}
\scriptsize
\setlength{\tabcolsep}{5.5pt}
\renewcommand{\arraystretch}{1.0}
\begin{tabular}{lcccccc}
\toprule
& \multicolumn{3}{c}{\textbf{AUROC}} & \multicolumn{3}{c}{\textbf{AUPRC }} \\
\cmidrule(lr){2-4}\cmidrule(lr){5-7}
\textbf{Drug} & \textbf{0.05} & \textbf{0.5} & \textbf{5.0} & \textbf{0.05} & \textbf{0.5} & \textbf{5.0} \\
\midrule
Fluorouracil
& \meanstdsmall{0.680}{0.172}$^{\dagger}$ & \underline{\meanstdsmall{0.705}{0.159}} & \textbf{\meanstdsmall{0.813}{0.153}}
& \meanstdsmall{0.749}{0.128}$^{\dagger}$ & \underline{\meanstdsmall{0.782}{0.120}} & \textbf{\meanstdsmall{0.831}{0.142}} \\
Buparlisib
& \underline{\meanstdsmall{0.544}{0.054}} & \textbf{\meanstdsmall{0.553}{0.044}} & \meanstdsmall{0.537}{0.055}$^{\dagger}$
& \textbf{\meanstdsmall{0.559}{0.056}} & \underline{\meanstdsmall{0.538}{0.061}} & \meanstdsmall{0.529}{0.051}$^{\dagger}$ \\
Alpelisib
& \meanstdsmall{0.631}{0.109}$^{\dagger}$ & \underline{\meanstdsmall{0.658}{0.064}} & \textbf{\meanstdsmall{0.678}{0.082}}
& \meanstdsmall{0.590}{0.089}$^{\dagger}$ & \underline{\meanstdsmall{0.591}{0.107}} & \textbf{\meanstdsmall{0.678}{0.066}} \\
Ruxolitinib
& \underline{\meanstdsmall{0.682}{0.221}} & \meanstdsmall{0.564}{0.144}$^{\dagger}$ & \textbf{\meanstdsmall{0.738}{0.182}}
& \textbf{\meanstdsmall{0.361}{0.268}} & \meanstdsmall{0.312}{0.245}$^{\dagger}$ & \underline{\meanstdsmall{0.351}{0.190}} \\
Ribociclib
& \textbf{\meanstdsmall{0.607}{0.072}} & \meanstdsmall{0.559}{0.074}$^{\dagger}$ & \underline{\meanstdsmall{0.572}{0.069}}
& \textbf{\meanstdsmall{0.535}{0.069}} & \meanstdsmall{0.501}{0.097}$^{\dagger}$ & \underline{\meanstdsmall{0.508}{0.069}} \\
Erlotinib
& \textbf{\meanstdsmall{0.767}{0.149}} & \underline{\meanstdsmall{0.733}{0.149}} & \meanstdsmall{0.600}{0.253}$^{\dagger}$
& \textbf{\meanstdsmall{0.776}{0.036}} & \underline{\meanstdsmall{0.759}{0.045}} & \meanstdsmall{0.608}{0.297}$^{\dagger}$ \\
Gemcitabine
& \textbf{\meanstdsmall{0.473}{0.334}} & \underline{\meanstdsmall{0.367}{0.290}} & \meanstdsmall{0.353}{0.233}$^{\dagger}$
& \textbf{\meanstdsmall{0.646}{0.254}} & \meanstdsmall{0.614}{0.232}$^{\dagger}$ & \underline{\meanstdsmall{0.625}{0.187}} \\
Paclitaxel
& \textbf{\meanstdsmall{0.634}{0.264}} & \meanstdsmall{0.523}{0.259}$^{\dagger}$ & \underline{\meanstdsmall{0.601}{0.183}}
& \textbf{\meanstdsmall{0.617}{0.294}} & \meanstdsmall{0.520}{0.251}$^{\dagger}$ & \underline{\meanstdsmall{0.535}{0.247}} \\
Tamoxifen
& \underline{\meanstdsmall{0.657}{0.211}} & \meanstdsmall{0.538}{0.296}$^{\dagger}$ & \textbf{\meanstdsmall{0.690}{0.163}}
& \textbf{\meanstdsmall{0.308}{0.387}} & \meanstdsmall{0.150}{0.070}$^{\dagger}$ & \underline{\meanstdsmall{0.183}{0.063}} \\
Trametinib
& \underline{\meanstdsmall{0.760}{0.189}} & \meanstdsmall{0.707}{0.243}$^{\dagger}$ & \textbf{\meanstdsmall{0.777}{0.209}}
& \textbf{\meanstdsmall{0.914}{0.058}} & \meanstdsmall{0.887}{0.083}$^{\dagger}$ & \underline{\meanstdsmall{0.909}{0.091}} \\
\bottomrule
\end{tabular}
\end{table*}

\section{Drug-specific Transition Utility and Reliability Analysis}
\label{app:transition_utility_reliability}

Aggregate target-domain evaluations assess whether the predicted transition provides additional response-prediction signal on average, but they do not establish whether this benefit is consistent across treatments. We therefore performed an exploratory, post-hoc drug-level analysis to characterize heterogeneity in the incremental utility of the transferred transition representation and to examine two potential explanations for this heterogeneity: the degree of representation of each target drug in the source perturbation atlas and properties of the corresponding target-domain response strata.

For each drug $d$, we define transition utility as the difference between the drug-specific performance of the transition-augmented response model and that of the corresponding static patient--drug baseline:
\begin{align}
\Delta \mathrm{AUROC}_d
&=
\mathrm{AUROC}_d(\mathrm{Patient+Drug+Transition})
-
\mathrm{AUROC}_d(\mathrm{Patient+Drug}),
\\
\Delta \mathrm{AUPRC}_d
&=
\mathrm{AUPRC}_d(\mathrm{Patient+Drug+Transition})
-
\mathrm{AUPRC}_d(\mathrm{Patient+Drug}).
\end{align}
All comparisons use the same target-domain splits and response-model procedures as the corresponding primary benchmark. These analyses are intended to characterize transfer behavior and identify potential reliability boundaries; they do not establish causal determinants of transition utility.

\subsection{Drug-specific transition benefit across target benchmarks}
\label{app:drug_specific_transition_benefit}

\begin{table*}[t]
\centering
\caption{
Drug-specific incremental utility of the transferred transition representation across target benchmarks. Values report mean AUROC across the corresponding outer folds or evaluation seeds. $\Delta$AUROC is defined as $\mathrm{AUROC}_{\mathrm{PerturbRx}}-\mathrm{AUROC}_{\mathrm{Patient+Drug}}$. Positive values indicate improved discrimination after adding the predicted transition.
}
\label{tab:drug_specific_transition_utility}
\small
\setlength{\tabcolsep}{5pt}
\begin{tabular}{llrrr}
\toprule
Benchmark & Drug & Patient+Drug & PerturbRx & $\Delta$AUROC \\
\midrule

\multirow{5}{*}{TCGA-508}
& Cisplatin    & 0.648 & 0.631 & -0.017 \\
& Docetaxel    & 0.636 & 0.711 & +0.075 \\
& Fluorouracil & 0.553 & 0.562 & +0.009 \\
& Gemcitabine  & 0.462 & 0.472 & +0.010 \\
& Paclitaxel   & 0.605 & 0.715 & \textbf{+0.110} \\
\midrule

\multirow{5}{*}{TCGA-186}
& Cisplatin     & 0.669 & 0.684 & +0.016 \\
& Fluorouracil  & 0.451 & 0.674 & \textbf{+0.223} \\
& Gemcitabine   & 0.378 & 0.551 & +0.173 \\
& Sorafenib     & 0.707 & 0.753 & +0.047 \\
& Temozolomide  & 0.651 & 0.653 & +0.002 \\
\midrule

\multirow{10}{*}{PDX}
& Alpelisib     & 0.664 & 0.631 & -0.033 \\
& Buparlisib    & 0.575 & 0.544 & -0.031 \\
& Erlotinib     & 0.567 & 0.767 & \textbf{+0.200} \\
& Fluorouracil  & 0.630 & 0.680 & +0.050 \\
& Gemcitabine   & 0.332 & 0.473 & +0.142 \\
& Paclitaxel    & 0.676 & 0.634 & -0.042 \\
& Ribociclib    & 0.609 & 0.607 & -0.002 \\
& Ruxolitinib   & 0.926 & 0.682 & \textbf{-0.243} \\
& Tamoxifen     & 0.871 & 0.657 & -0.214 \\
& Trametinib    & 0.737 & 0.760 & +0.023 \\
\bottomrule
\end{tabular}
\end{table*}

Transition utility varied substantially across drugs and target benchmarks. On TCGA-508, the largest AUROC gains were observed for Paclitaxel ($\Delta\mathrm{AUROC}=+0.110$) and Docetaxel ($+0.075$), whereas the changes for Gemcitabine ($+0.010$) and Fluorouracil ($+0.009$) were small and Cisplatin showed a slight decrease ($-0.017$).

On TCGA-186, the transition-augmented representation improved drug-specific AUROC for all five evaluated drugs. The largest gains were observed for Fluorouracil ($+0.223$) and Gemcitabine ($+0.173$), followed by Sorafenib ($+0.047$), Cisplatin ($+0.016$), and Temozolomide ($+0.002$).

The PDX benchmark showed greater drug-level heterogeneity. Erlotinib ($+0.200$), Gemcitabine ($+0.142$), Fluorouracil ($+0.050$), and Trametinib ($+0.023$) showed improved AUROC, whereas several drugs showed little change or decreased performance. The largest decreases occurred for Ruxolitinib ($-0.243$) and Tamoxifen ($-0.214$).

Overall, these results indicate that the transferred transition is not uniformly beneficial across treatments. We therefore next examined whether this heterogeneity was associated with the degree of source-atlas representation of each target drug.

\subsection{Association with source-drug chemical similarity}
\label{app:transition_chemical_similarity}

\begin{table}[b]
\centering
\caption{
Rank associations between drug-specific transition utility, measured by $\Delta$AUROC, and candidate explanatory factors. For chemical similarity, $p$-values are based on two-sided exact permutation tests. PDX target-domain support variables use the same exact permutation framework over the ten-drug PDX benchmark.
}
\label{tab:transition_utility_associations}
\small
\setlength{\tabcolsep}{6pt}
\begin{tabular}{llrr}
\toprule
Benchmark & Factor & $\rho$ & $p_{\mathrm{exact}}$ \\
\midrule

\multirow{2}{*}{TCGA-508}
& Morgan ECFP4 Tanimoto
    & +0.900 & 0.083 \\
& ChemBERTa cosine
    & +0.900 & 0.083 \\
\midrule

\multirow{2}{*}{TCGA-186}
& Morgan ECFP4 Tanimoto
    & +0.100 & 0.950 \\
& ChemBERTa cosine
    & +0.500 & 0.450 \\
\midrule

\multirow{6}{*}{PDX}
& Morgan ECFP4 Tanimoto
    & -0.067 & 0.865 \\
& ChemBERTa cosine
    & -0.030 & 0.946 \\
& Total sample size
    & -0.527 & 0.123 \\
& Minority-class size
    & +0.134 & 0.707 \\
& Minority-class fraction
    & +0.370 & 0.296 \\
& Responder fraction
    & +0.673 & 0.039 \\
\bottomrule
\end{tabular}
\end{table}

We first asked whether target drugs that are more closely represented by compounds in the source perturbation atlas tend to derive greater benefit from the transferred transition. For each target drug, we quantified nearest-source chemical similarity using two complementary molecular representations: Morgan ECFP4 fingerprint Tanimoto similarity and cosine similarity between ChemBERTa embeddings. We then evaluated the Spearman rank association between nearest-source similarity and drug-specific transition utility.

On TCGA-508, both similarity measures showed a large positive rank association with $\Delta$AUROC ($\rho=0.90$ for both Morgan and ChemBERTa). However, because this analysis contains only five drugs, the corresponding two-sided exact permutation tests did not reach conventional significance ($p_{\mathrm{exact}}=0.083$ for both measures). The corresponding association with $\Delta$AUPRC was weaker ($\rho=0.70$, $p_{\mathrm{exact}}=0.233$).

This pattern did not replicate consistently across the other target benchmarks. On TCGA-186, the AUROC associations were $\rho=0.10$ for Morgan similarity ($p_{\mathrm{exact}}=0.950$) and $\rho=0.50$ for ChemBERTa similarity ($p_{\mathrm{exact}}=0.450$). In the larger 10-drug PDX benchmark, the corresponding associations were close to zero: $\rho=-0.067$ for Morgan similarity ($p_{\mathrm{exact}}=0.865$) and $\rho=-0.030$ for ChemBERTa similarity ($p_{\mathrm{exact}}=0.946$). AUPRC gain was similarly not associated with either chemical-similarity measure in PDX ($\rho=-0.006$, $p_{\mathrm{exact}}=1.000$ for Morgan; $\rho=0.055$, $p_{\mathrm{exact}}=0.892$ for ChemBERTa).

Thus, although TCGA-508 showed an exploratory positive trend, nearest-source chemical similarity did not provide a consistent explanation for drug-specific transition utility across target benchmarks. In particular, the near-zero associations observed in PDX indicate that high source-drug chemical similarity alone is neither sufficient nor generally predictive of downstream transition benefit.

\subsection{Response composition and low-response regimes in PDX}
\label{app:pdx_transition_reliability}

Because source-drug chemical similarity did not consistently explain the heterogeneity in PDX transition utility, we next examined whether drug-specific benefit was associated with characteristics of the target-domain response distribution. We considered total sample size, minority-class size, minority-class fraction, and observed responder fraction for each drug.

Total sample size was not clearly associated with $\Delta$AUROC ($\rho=-0.527$, $p_{\mathrm{exact}}=0.123$), nor was minority-class size ($\rho=0.134$, $p_{\mathrm{exact}}=0.707$) or minority-class fraction ($\rho=0.370$, $p_{\mathrm{exact}}=0.296$). In contrast, the observed responder fraction showed a positive rank association with transition utility ($\rho=0.673$, $p_{\mathrm{exact}}=0.039$).

\begin{figure}[b]
    \centering
    \includegraphics[
        width=0.90\columnwidth
    ]{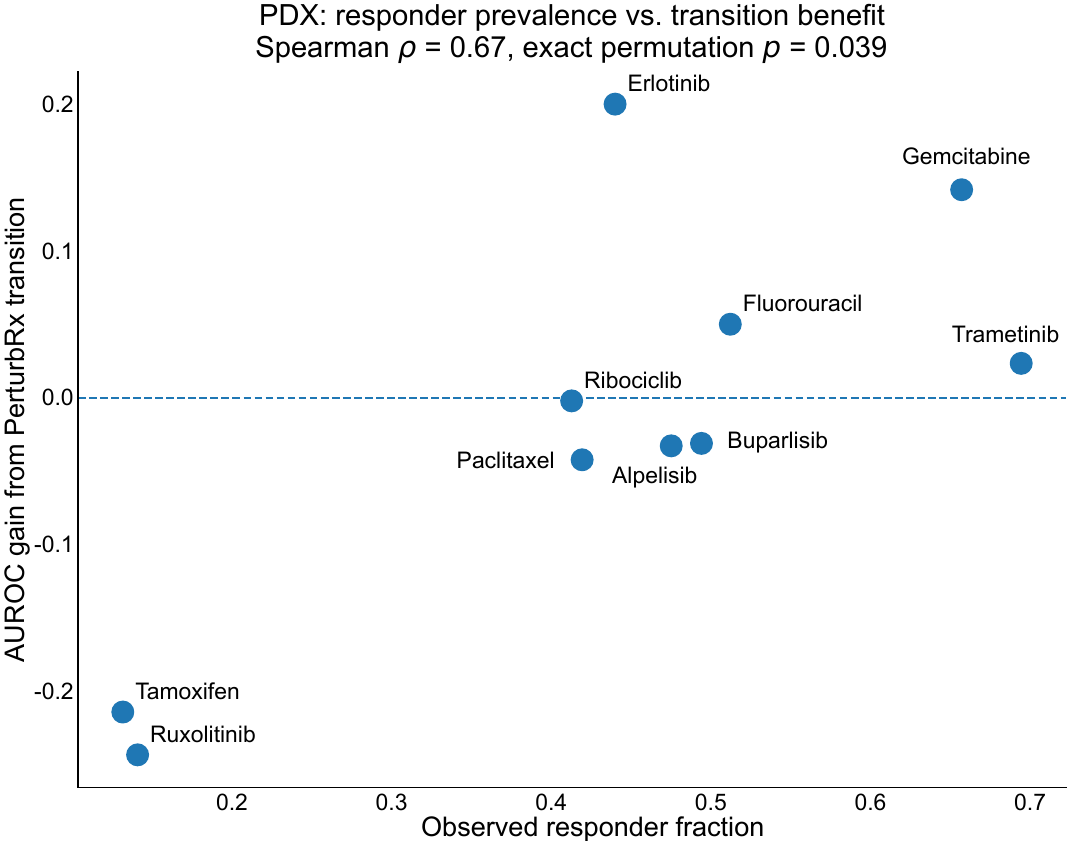}
    \caption{
    \textbf{PDX responder prevalence and drug-specific transition utility.}
    Each point represents one drug in the shared 10-drug PDX benchmark.
    Transition utility is measured as the change in drug-specific AUROC after
    augmenting the static patient--drug representation with the predicted
    transition. The observed responder fraction was positively rank-associated
    with transition utility ($\rho=0.673$, two-sided exact permutation
    $p=0.039$). The two largest decreases occurred for Tamoxifen and
    Ruxolitinib, which also had the lowest observed responder fractions.
    }
    \label{fig:pdx_responder_transition_benefit}
\end{figure}

This association should be interpreted cautiously. The two largest AUROC decreases occurred for Tamoxifen and Ruxolitinib, which also had the two lowest observed responder fractions in the PDX benchmark: 5 of 38 observations (13.2\%) for Tamoxifen and 10 of 71 (14.1\%) for Ruxolitinib. Leave-one-drug-out analyses preserved a positive association in every case, with $\rho$ ranging from 0.57 to 0.83. However, jointly excluding Tamoxifen and Ruxolitinib reduced the association to $\rho=0.381$ ($p=0.352$), indicating that the observed trend is driven in part by these two low-response strata.

\end{document}